\documentclass{JHEP3}
\usepackage{graphicx,amsmath,amssymb}
\usepackage{epsfig,multicol}
\usepackage{epsf}
\usepackage{epstopdf}
\input{epsf.sty}
\newcommand{\ba}{\begin{eqnarray}}
\newcommand{\ea}{\end{eqnarray}}
\newcommand{\D}{\overline{\mbox{D}}}
\newcommand{\e}{\epsilon}

\def\beq{\begin{equation}}
\def\eeq{\end{equation}}
\def\beqa{\begin{eqnarray}}
\def\eeqa{\end{eqnarray}}
\def\ba{\begin{eqnarray}}
\def\ea{\end{eqnarray}}
\def\be{\begin{equation}}
\def\ee{\end{equation}}

\def\hf{\frac{1}{2}}
\def\pprime{\prime\prime}
\def\ap{\alpha^{\prime}}

\def\D{\bar D}

\title{Observing Brane Inflation}
\author{Sarah~E. Shandera\footnote{seb56@mail.lns.cornell.edu}  \:  and S.-H. Henry Tye 
\footnote{tye@lepp.cornell.edu}
\\ Laboratory for Elementary Particle Physics \\ 
Cornell University\\ 
  Ithaca, NY 14853}

\abstract{
Linking the slow-roll scenario and the Dirac-Born-Infeld scenario of ultra-relativistic roll (where, thanks to the warp factor, the inflaton moves slowly even with an ultra-relativistic Lorentz factor), we find that the KKLMMT $D$3-${\D}$3-brane inflation is robust, that is, enough e-folds of inflation is quite generic in the parameter space of the model. We show that the intermediate regime of relativistic roll can be quite interesting observationally. Introducing appropriate inflationary parameters, we explore the parameter space and give the constraints and predictions for the cosmological observables in this scenario.
Among other properties, this scenario allows the saturation of the present observational bound of 
either the tensor/scalar ratio $r$ (in the intermediate regime) or the non-Gaussianity $f_{NL}$ (in the ultra-relativistic regime), but not both.}

\begin{document}

\maketitle

\section{Introduction}

The inflationary universe was proposed to solve a number of fine-tuning problems such as the flatness problem, the horizon problem and the defect problem  \cite{Guth:1980zm}. Its prediction of an almost scale-invariant density perturbation power spectrum has received strong observational support from the temperature fluctuation in the cosmic microwave background radiation \cite{Smoot:1992td,WMAP}. 
It is believed by many that superstring theory is the fundamental theory of all matter and forces, so it is natural to look there for an explicit realization of the inflationary scenario. A simple, realistic and well-motivated model is the $D$3-${\D}$3-brane 
inflation \cite{Dvali:1998pa,collection} where the 6 compactified dimensions are dynamically stabilized \cite{Giddings:2001yu,Kachru:2003aw}. This is the KKLMMT scenario \cite{Kachru:2003sx}.
Here, inflation ends when the $D$3-brane and the ${\D}$3-brane collide and annihilate, initiating 
the hot big bang epoch. The annihilation of the $D$3-${\D}$3-branes allows the universe to settle down to the string vacuum state that describes our universe. During the inflationary epoch, the ${\D}$3-brane
is expected to sit at the bottom of a warped throat (due to an attractive force) while the $D$3-brane is relatively mobile. It is possible (in fact one may argue likely) that the inflaton potential has relatively flat directions outside the throat, allowing substantial inflation. Unfortunately, the precise potential is rather dependent on the detailed structures of the compactification and remains to be explored more carefully. To avoid this issue, we shall assume here that the $D$3-brane starts close to or inside the throat. If we have enough e-folds in the throat, then the physics outside the throat need not concern us. That is, we are looking for the sufficient condition of enough e-folds, and this condition will be relaxed if some inflation has already taken place before the $D$3-brane reaches the edge of the throat. 
Warped throats such as the Klebanov-Strassler (KS) conifold construction \cite{Klebanov:2000hb} are generic in any flux compactification that stabilizes the moduli. The Dirac-Born-Infeld (DBI) action for the inflaton field follows simply because the inflaton is an open string mode. 
By now it is clear that enough inflation is generic in this scenario thanks to : \\
(1) the warped geometry of the throat in a realistic string compactification, which tends to flatten (by orders of magnitude) the attractive Coulombic potential between the $D$3-brane and the $\D$3-brane \cite{Kachru:2003sx}; \\
(2) the warped geometry of the throat combined with the DBI action, which forces the inflaton to move slowly as it falls towards the bottom of the throat, as pointed out by Silverstein and Tong \cite{Silverstein:2003hf} and with Alishahiha \cite{Alishahiha:2004eh}. \\
Let us elaborate on these points.

Close to and inside the throat, the potential takes the simple approximate form
\be
\label{potential}
V(\phi) = V_K(\phi) + V_0 +V_C(\phi) \simeq \frac{m^2}{2}\phi^2 + V_{0}\left(1-\frac{vV_{0}}{4\pi^{2}}\frac{1}{\phi^4}\right) 
\ee
where the constant term $$V_{0}=2T_3h_A^4= 2 T_{3} h(\phi_{A})^{4}$$
is the effective cosmological constant.
Here $\phi$ is the canonical inflaton ($\phi= \sqrt{T_{3}} r$, where $r$ is the position of the $D$3 brane with respect to the bottom of the throat), $T_{3}$ is the $D$3-brane tension and $h_A$ is the warp factor $h(\phi)$ at $\phi_A$, the position of the ${\D}$3-brane. The factor $v$ depends on the properties of the warped throat, with $v=27/16$ for the KS throat.
With some warping (say, $h_{A} \simeq 1/5$ to $10^{-3}$), the attractive Coulombic potential 
$V_C(\phi)$ can be very weak (i.e., flat).
The quadratic term $V_K(\phi)$ receives contributions from a number of sources and is rather model-dependent \cite{Kachru:2003sx,Shandera:2004zy}. However $m^{2}$ is expected to be comparable to $H_{0}^2=V_{0}/3M_{p}^{2}$, where $M_{p}$ is the reduced Planck mass ($G^{-1}=8 \pi M_{p}^{2}$). This sets the canonical value for the inflaton mass $m_{0}=H_{0}$ (which turns out to be around $10^{-7}M_{p}$). 

For a generic value of $m$, usual slow-roll inflation will not yield enough e-folds of inflation. Ref.\cite{Firouzjahi:2005dh} shows that $m \lesssim m_{0}/3$ will be needed. Na\"{i}vely, a substantially larger $m$ will be disastrous, since the inflaton will roll fast, resulting in very few e-folds in this case. However, for a fast roll inflaton, string theory dictates that we must include higher powers of the time derivative of $\phi$, in the form of the DBI action
\be
S=-\int d^4x\;a^3(t)\left[T\sqrt{1- \dot{\phi}^2/T} + V(\phi) - T \right]
\ee
where $T(\phi) = T_{3}h(\phi)^{4}$ is the warped $D$3-brane tension at $\phi$. For the usual slow-roll,
$T\sqrt{1- \dot{\phi}^2/T}  - T \simeq  \dot{\phi}^2/2$.
It is quite amazing that the DBI action now allows enough e-folds even when the inflaton potential is steep \cite{Silverstein:2003hf,Alishahiha:2004eh}. To see why this happens, consider the Lorentz factor $\gamma$ from the DBI action,
\be
\label{gammabound}
\gamma = \frac{1}{\sqrt{1- \dot{\phi}^2/T(\phi)}} \quad \rightarrow \quad \dot{\phi}^2<T(\phi)
\ee
where $T(\phi)=T_{3}h(\phi)^{4} \simeq \phi^{4}/\lambda$ during inflation. As the $D$3-brane approaches $\D$3-brane, $\phi$ and $T(\phi)$ decrease, and $h(\phi) \rightarrow h(\phi_{A})$. 
The key is that $\dot{\phi}$ is bounded by Eq.(\ref{gammabound}), and this bound gets tighter as $T(\phi)$ decreases. This happens even if the potential is steep, for example, when $m > H_{0}$. 
So the inflaton rolls slowly either because the potential is relatively flat (so $\gamma \simeq1$ in the usual slow-roll case), or because the warped tension $T(\phi)$ is small (so $1 \ll \gamma < \infty $).  
As a result, it can take many e-folds for $\phi$ to reach the bottom of the throat. 
This implies that the brane inflationary scenario is very robust.
When $\gamma \gg 1$, the kinetic energy is enhanced by a Lorentz factor of $\gamma$. Note that the inflaton is actually moving slowly down the throat even in this relativistic limit.
However, the characteristics of this scenario are very different from the usual slow-roll limit, where $\gamma \simeq 1$.
To draw a distinction, we call this the ultra-relativistic regime. Unless otherwise specified, $\gamma$ below refers to $\gamma_{55}$, i.e., $\gamma$ at 55 e-folds before the end of inflation.

In this paper, we would like to find the observational constraints on the general properties of this scenario. We present a more complete analysis of the ultra-relativistic scenario than that in Ref.\cite{Alishahiha:2004eh}. We also extend the analysis to the intermediate (relativistic) region ($\gamma > 1$), i.e., intermediate values of $m$. Not surprisingly, the analysis of the intermediate region is somewhat more involved than either the slow-roll (i.e., non-relativistic) region \cite{Firouzjahi:2005dh} 
or the ultra-relativistic region \cite{Alishahiha:2004eh}. To see this, one can do a simple counting of parameters. From the potential, we have 3 parameters, namely, $m$, $\lambda$ and $h_A$. We also have two constraints: we must match the COBE normalization for the density perturbations and we must obtain at least 55 e-folds in the throat. Unfortunately, these constraints do not fix the parameter values except in special cases (e.g., slow-roll) because implementing them requires us to also specify the endpoints of inflation, $\phi_i$ and $\phi_f$. $\phi_f$, the end of inflation, is taken care of by the position of the anti-brane, which is related to $h_A$, so this is not a new parameter. However, the starting point of inflation and the condition that the $D$3-brane should be inside the throat is important, particularly in non-slow-roll cases. We clarify with a few points:

$\bullet$ In the usual slow-roll case, there are essentially 2 parameters : $m$ and $V_{0}$. After fitting the COBE density perturbation data  \cite{Smoot:1992td}, the predictions are reduced to a one parameter analysis \cite{Firouzjahi:2005dh}. The parameter $\phi_{end}$, the value when the tachyon mode appears, signals the collision and annihilation of the branes. Since $\phi_{end}$ is close to the final value $\phi_{f}$ (when inflation ends) and $\phi_{A}$ (the bottom of the throat), these differences may be ignored. The constraint  that the $D$3-brane should be inside the throat at 55 e-folds is rather weak here. This can be seen from the uniformity along the throat: growth in the Lorentz factor $\gamma$ depends on the derivative of the Hubble parameter. For a nearly constant potential, the Hubble parameter and $\gamma$ are also very nearly constant, so changing $\phi_i$ has little significance.\\
$\bullet$ In the ultra-relativistic case, $m$ is large so $V_{K}$ dominates (i.e., $V_{0}$ can be ignored), so the problem is again reduced to the above 3 parameters before imposing the COBE normalization. In this scenario, ensuring that all 55 e-folds of inflation take place while the $D$3-brane is inside the throat becomes a strong constraint; that is, the ``initial'' position $\phi_{i}$ (at 55 e-folds before the end of inflation) should satisfy  $\phi_{i} \le \phi_{e}$ where $\phi_{e}$ is the value at the edge of the throat, i.e., $h(\phi_{e}) \simeq 1$. Also, shifting $\phi_i$ can shift $\gamma$ at 55 e-folds, which also significantly shifts the observables. To implement the condition that inflation happens entirely in the throat, we need to introduce the $D$3-brane tension $T_{3}$, or the string scale $\ap$, beyond its appearance in  the combination $T$.
Since $V_{0}$ can be ignored in this case, one may obtain all the inflationary properties without the $\D$3-brane. However, one should check if  ``graceful exit'' (reheating or preheating) can be successfully realized in such a scenario. \\
$\bullet$ In the intermediate region, in addition to the above parameters and the constraint, $V_{0}$ and the Coulomb potential $V_{C}(\phi)$ comes into play. As $m$ increases and becomes bigger than $H_{0}$, the slow-roll approximation breaks down. As pointed out earlier, the presence of both terms and the constraint that the $D$3-brane should be inside the throat during inflation brings in the dependence of all the parameters of the theory, namely, $m$, $V_{0}$, $\ap$, $h_{A}$ and the string coupling $g_{s}$. To simplify the analysis, we fix $g_{s} \simeq 0.1$ throughout. Since the only data used is the COBE normalization, variations in the input parameters lead to a range of predictions for the other CMBR numbers.

To study the general DBI brane inflationary scenario, it is convenient to introduce a new set of inflationary parameters, namely, $\eta_{D}$, $\epsilon_{D}$ and $\kappa_{D}$, generalizing the usual slow-roll parameters.  As we shall see, $\lambda$, $h_{A}$ and $V_{0}$ will be related to other string theory parameters as well as $\ap$. 
 
\begin{figure}[htb]
\begin{center}
\includegraphics[width=0.6\textwidth,angle=0]{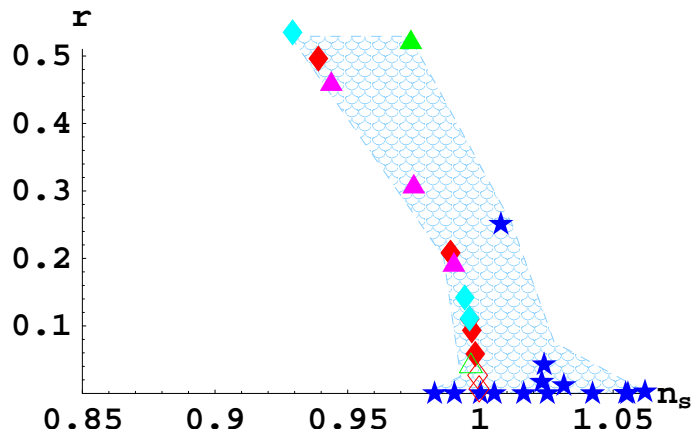} 
\caption{The tensor/scalar ratio $r$ versus the scalar power index $n_{s}$. The shaded region is covered by the model. A few sample points are shown: stars are slow-roll points, triangles and diamonds are relativistic. Most of the region is filled out by relativistic points, where higher $m^2$ in the potential and/or large $\gamma$ lead to points further to the left (smaller $n_s$). The outlined (i.e., open) symbols have $\gamma$ too high to satisfy the current non-Gaussianity bound. To understand the slope of the left boundary, note that large $r$ requires a steeper potential (out of the slow-roll regime), which also corresponds to $n_s<1$. 
\label{rnsplot}}
\end{center}
\end{figure}

\begin{figure}[htb]
\begin{center}
\includegraphics[width=0.6\textwidth,angle=0]{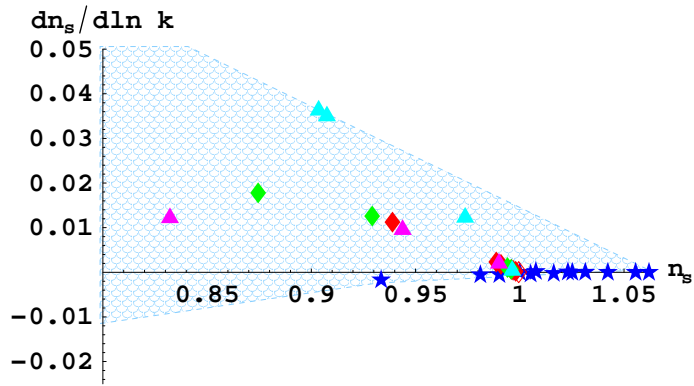} 
\caption{The running of the scalar index $\frac{d n_{s}}{d \ln k}$ versus the scalar index $n_{s}$. The shaded region is covered by our model. A few sample points are shown: stars are slow-roll points, triangles and diamonds are relativistic. Most of the region is filled out by relativistic points, where higher $m^2$ in the potential and/or large $\gamma$ lead to points further to the left (smaller $n_s$). The outlined symbols have $\gamma$ too high to satisfy the current non-Gaussianity bound. 
\label{nsdnsplot}}
\end{center}
\end{figure}

\begin{figure}[htb]
\begin{center}
\includegraphics[width=0.6\textwidth,angle=0]{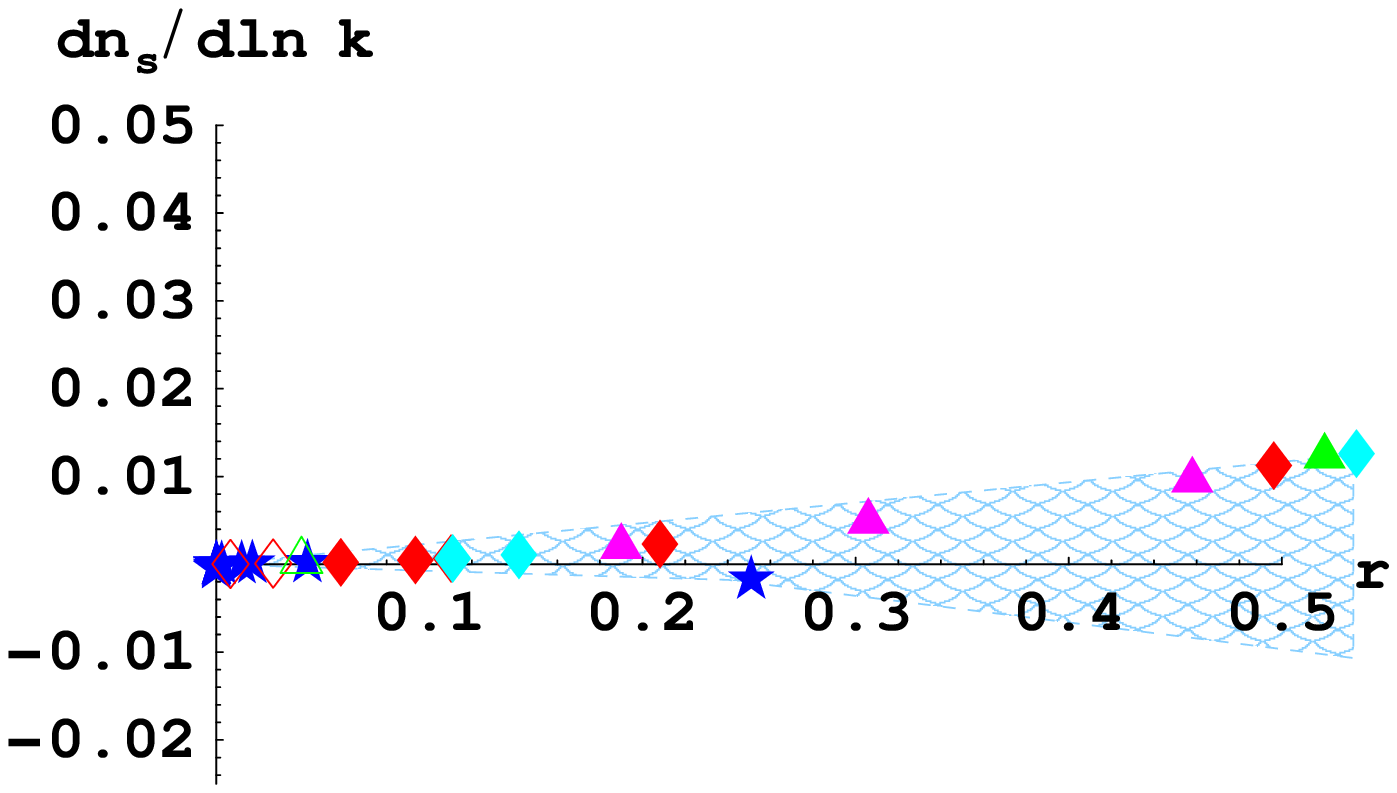} 
\caption{The running of the scalar index $\frac{d n_{s}}{d \ln k}$ versus the tensor/scalar ratio $r$. The shaded region is covered by our model. A few sample points are shown: stars are slow-roll points, triangles and diamonds are relativistic. Most of the region is filled out by relativistic points, where higher $m^2$ in the potential and/or large $\gamma$ fills in points further to the right (larger $r$). The outlined symbols have $\gamma$ too high to satisfy the current non-Gaussianity bound. 
\label{rdnsplot}}
\end{center}
\end{figure}

Here we summarize the key predictions for the cosmological observables, namely, the scalar power index $n_{s}$, its running $\frac{d n_{s}}{d \ln k}$, the tensor to scalar ratio $r$, the non-Gaussianity  $f_{NL}$ and the cosmic string tension $\mu$ (since cosmic strings are generically produced towards the end of brane inflation \cite{cosmicstring}). We use a simplified warp factor, so the predictions should be treated as semi-quantitative. Figures \ref{rnsplot}, \ref{nsdnsplot} and \ref{rdnsplot} show the correlations between different observables in the model. The points come from varying the parameters in the potential (and the background geometry) to find at least 55 e-folds in the throat, and ensuring that the density perturbation generated at $N_{e}=55$ corresponds to that measured by COBE \cite{Smoot:1992td}.  

Each figure shows a shaded region indicating the values obtainable in this model. We also show a few sample points. Stars (blue) are slow-roll points calculated for various values $m$. These cluster at $n_s\sim1$, small $r$ and small scalar running as expected. The diamonds and triangles show trajectories of increasing $\gamma$ for particular $m$ values (differentiated by color).  As $\gamma$ increases, the observables return toward an exactly scale-invariant spectrum with $r=0$. Large $\gamma$ leads to large non-Gaussianities, so outline points (open diamonds and triangles) show scenarios that exceed that observational non-Gaussianity bound. Some points do not appear in a particular graph because they take values outside the given ranges. \\

A few comments are in order here :

$\bullet$ There is a parameter space region beyond the slow-roll region where the ratio $r$ saturates the present WMAP observational bound. This region corresponds to relativistic ($1.2 \lesssim \gamma\lesssim 6$) roll, $n_s<1$, and relatively large scalar running. For example, the diamond point with largest $r$ in Fig.(\ref{rnsplot}) has $\gamma=2.4$, $n_s=0.93$, and $\frac{d n_{s}}{d\ln k}=10^{-2}$. It comes from the parameter choice $m/M_p=6\times10^{-5}$, $N=2.5\times10^{12}$, $h_A=0.003$, and $\ap=1600/M_p^2$.

$\bullet$ Even if the brane is slow-rolling at $N_{e} \sim 55$, $\gamma$ can end up large ($\gamma \gg 1$) as the brane drops down the throat. This changes the estimate of the number of e-folds and so modifies the naive predictions of the inflationary parameters. Without the DBI action, the naive number of e-folds will be smaller. Crudely, $N_{e} \propto \gamma$. 

$\bullet$ We find that it is possible to keep $\gamma$ small ($\gamma < 2$) for the full range of the quadratic term investigated ($0 \le m<10^{-4}M_p$). For $\gamma$ not too big ($<6$), the tensor/scalar ratio $r$ can saturate the current bound ($r\lesssim0.5$). 

$\bullet$ It was pointed out in Ref.(\cite{Alishahiha:2004eh}) that the non-Gaussianity $f_{NL}$ is proportional to $\gamma^2$. The present observational bound from non-Gaussianity \footnote{The numerical value of the coefficient of $\gamma^2$ in $f_{NL}$ was corrected from previously reported values by X. Chen, M. Huang and G. Shiu (personal communication).} $|f_{NL}|  \simeq 0.32 \gamma^{2} \lesssim 300$ yields $\gamma \lesssim 31$, where the limit on non-Gaussianities coming from DBI-type models is discussed in Ref.\cite{Creminelli:2005hu}. However, in the intermediate regime, increasing $N$ can be thought of as decreasing the value of $\phi$ where the correct COBE normalization is found. Moving lower in the throat corresponds to larger $\gamma$, and we find it possible to achieve $\gamma \gg 31$. To keep this consistent with 55-efolds, $h_A$ must be decreased (or alternatively $\ap$ decreased, which shifts the edge of the throat and the scaling of $\phi$ without changing the ratio of background fluxes). By increasing $\ap$ and adjusting the $\lambda$ and $h_A$ to maintain the COBE normalization at 55 e-folds, we find that it is possible to achieve $\gamma \gg 31$. For example, for $m/M_p=6\times10^{-5}$ and $\ap=1600/M_p^2$, $N=5\times10^{11}$ gives $\gamma=1.09$ and $N=3.02\times10^{12}$ gives $\gamma=11$. We cannot achieve higher $\gamma$ and still find 55 e-folds in the throat except by increasing $\ap$. With $\ap=10^4/M_p^2$ and $N=3.09\times10^{12}$, we find $\gamma=194$. In all cases, as $N$ increases, $h_A$ must be lowered to keep the correct COBE normalization at 55 e-folds. As soon as $\gamma$ is larger than about 1.05, significant deviations from the expected slow-roll behavior show up.
 
$\bullet$ Cosmic strings are generically produced toward the end of brane inflation. We find that the cosmic string tension $\mu$ roughly satisfies $10^{-14}<G\mu<10^{-6}$; its particular value depends on the choice of $m$ and $\lambda$. Roughly speaking, small allowed values of $G\mu$, $10^{-14} \lesssim G\mu \lesssim 10^{-10}$ are accompanied by a red tilt in the power spectrum, and only this range of tensions can be accompanied by large non-Gaussianity.

The predictions for the inflationary properties are quite different as we vary $m$. We find it useful to consider the following regions : $m \ll H_{0}$ (the KKLMMT scenario \cite{Kachru:2003sx} for slow-roll inflation \cite{Firouzjahi:2005dh}), $m \gg H_{0}$, and the intermediate region $m \sim H_{0}$. For  $m \gg H_{0}$, the small $m$ and small $H_0$ region reduces to the well-known chaotic inflation with slow-roll \cite{Linde:1988dk}. 
Otherwise, it is in the ultra-relativistic region \cite{Silverstein:2003hf,Alishahiha:2004eh} or the relativistic region. In this paper, we consider the whole reasonable range of $m$ by extending previous analyses to the region where $m \sim H_{0}$. 
We may also consider $D$3/$D$7-brane inflation \cite{Dasgupta:2002ew}, where again the $D$3-brane position is the inflaton. 
For tachyonic inflaton mass ($m^{2} <0$), the scenario becomes the multi-throat brane inflation scenario proposed by Chen \cite{Chen:2004gc}.

In the rest of the paper, we give a brief review and set up the procedure that allows us to treat the full mass range (\S2), review and combine the usual brane inflation scenarios (\S3), analyze the regime between slow-roll and DBI inflation (\S4) and discuss our results (\S5).

\section{The Model}

The setup for the $D$3-$\bar{D}3$-brane inflation in string theory was given in Ref.\cite{Kachru:2003sx}. The essential elements are IIB string theory compactified on a Calabi-Yau manifold with one or more warped throats. Inflation may occur either when a $D$3-brane moves toward some $D$7-branes wrapping some 4-cycles, or between a $D$3-brane and a $\D$3-brane where the antibrane sits at the bottom of a warped throat. These two models share many features, but we will work in the $D$3-$\D$3 case. Although we are concerned with a general model that need not be slow-roll or ultra-relativistic, much of this section follows closely the derivations in \cite{Silverstein:2003hf}.

We start with the metric 
\be
ds_{10}^{2} = h^{2}(r)[-dt^{2} +a(t)^{2}dx^{2}]  + h^{-2}(r) ds_{6}^{2}
\ee
where $h$ is the (dimensionless) warp factor, which depends on the properties of the throat. We are interested in warped conifolds with fluxes, described by non-compact Calabi-Yau manifolds. The best known example is the warped deformed conifold of $T^{1,1}$ \cite{Klebanov:2000hb}.
To simplify the analysis, we shall mostly consider the following simplified $h$, 
\be
\label{warpfunction}
h^{-4}(\phi) \simeq \frac{T_3\lambda}{\phi^4}
\ee
where $\lambda\equiv T_3R^4$ with
\ba
R^{4} = 4 \pi g_s N  {\alpha^{\prime}}^{2} v 
\ea
and $v$ is the ratio of the volume of the base manifold with respect to that of the unit $S^5$ in $AdS_{5} \times S^{5}$. For $T^{1,1}$, $v=27/16$.

In this metric, the DBI action for a $D$3-brane becomes
\be
\label{action}
S=-T_3\int d^4x\;a^3(t)\left[h^4\sqrt{1-h^{-4}\dot{\phi}^2T_3^{-1}}+T_3^{-1}V(\phi)- h^4\right]
\ee
where the radial component $r$ of the gravity coordinate ${\bf r}$ of the $D$3-brane becomes the field theory inflaton $\phi=\sqrt{T_3}r$. The $\bar{D}$3-brane is sitting at the bottom of the throat at $\phi=\phi_{A}$. 
Using the relations
\ba
T_{3} &=& \frac{1}{ (2\pi)^3 g_{s } {\alpha^{\prime}}^{2}} \nonumber \\
V_{0} &= & 2T_3h_A^4 = \frac{4 \pi^{2} \phi_{A}^{4}}{vN}=\frac{2 \phi_{A}^{4}}{\lambda}
\ea
we may write the 4-dim. effective inflaton potential (\ref{potential}) in the following form
\be
\label{potential1}
V  = \frac{m^2}{2}\phi^2 +V_{0}\left(1-\frac{vV_{0}}{4 \pi^{2}\phi^4}\frac{(\gamma+1)^2}{4\gamma}\right)
\ee
where $\gamma (\phi)$ is the ''Lorentz'' factor that allows for a fast-moving brane (see Appendix A). 
For slow roll, $\gamma\simeq 1$.
For ultra-relativistic roll ($\gamma \gg 1 $), the Coulombic term is negligible, so the last factor involving $\gamma$ is relevant only for the intermediate ($\gamma > 1 $) region.
We make use of this potential (\ref{potential1}) with the DBI action to obtain an equation of motion for the inflaton. We note that the model has 5 parameters, namely, $g_{s}$, $\alpha^{\prime}$, $m$, $V_{0}$ and $\lambda$. Even if we ignore $g_{s}$ (that is, fix it to $g_{s}=1$ or $1/10$), we still have 4 parameters, which leads to an interesting interpolation between the two end regimes where we have good analytic control.

From the action given in Eq.(\ref{action}), the pressure $p$ and density $\rho$ are given by\footnote{The $\gamma$ factors in the $D\bar{D}$ term give a correction to these equations which is suppressed by $1/N$, so we ignore it.}
\ba
\label{prho}
\rho=&T_3h^4(\gamma-1)+V \\\nonumber
p=&T_3h^4(\gamma - 1)/\gamma - V
\ea
where $\gamma$ is the Lorentz factor
$1/\gamma=\sqrt{1-\dot{\phi}^2/T_3 h(\phi)^{4}}$.
Notice that for small velocity, Eq.(\ref{prho}) reduces to the usual expressions. 
The Friedmann equations are
\ba
\label{Friedmann}
3H^2&=&\frac{1}{M_p^2}\rho\\\nonumber
2\frac{\ddot{a}}{a}+H^2&=&-\frac{1}{M_p^2}p
\ea
The equation of motion for the inflaton is
\be
\ddot{\phi}-\frac{6h^{\prime}}{h}\dot{\phi}^2+4T_3h^3h^{\prime}+\frac{3H}{\gamma^2}\dot{\phi}+\left(V^{\prime}-4T_3h^3h^{\prime}\right)\frac{1}{\gamma^3}=0
\ee
where $\gamma$, $H$, and $V$ are all functions of $\phi$ and the prime denotes derivatives with respect to $\phi$. We have used the continuity equation
\be
\dot{\rho}=-3H(\rho+p).
\ee
Some algebra gives a useful expression for $\dot{\phi}$
\be
\label{phidot}
\dot{\phi}=\frac{-2H^{\prime}}{\sqrt{1/M_p^4+4h^{-4}T_3^{-1}H^{\prime2}}}
\ee
Eq.(\ref{Friedmann}) means we need $w=p/\rho<-1/3$. 
Using Eq.(\ref{prho}), we have 
\be
\frac{\ddot{a}}{a} = \frac{V}{3M_{p}^{2}} - \frac{T(\gamma +2 -3/\gamma)}{6M_{p}^{2}}
\ee
where the kinetic term always contributes negatively. 

Since $\phi$ decreases monotonically with time $t$, it is convenient to use $\phi$ as the time variable.
We can now express everything in terms of $H(\phi)$. The most useful equation allows us to find $H(\phi)$ given the form of the potential. This equation is not necessarily easy to solve, but it can be done analytically in some cases and numerically for the regime between slow-roll and the ultra-relativistic case. Using Eq.(\ref{prho}) and Eq.(\ref{Friedmann}), we find
\ba
\label{solve}
V(\phi)&=&3M_p^2H(\phi)^2-T_3h(\phi)^4(\gamma(\phi)-1)\\\nonumber
\gamma(\phi) &=& \sqrt{1+4M_p^4T_3^{-1}h(\phi)^{-4}H^{\prime}(\phi)^2}\\\nonumber
\dot{\phi}(\phi)&=&\frac{-2M_p^2H^{\prime}(\phi)}{\gamma(\phi)}
\ea
From the first line, it is clear that the warp factor will be more significant for large $\gamma$. Combining the first and second lines gives the differential equation for $H(\phi)$.

Since we do not limit ourselves to slow-roll, we need parameters other than the usual slow-roll parameters to describe the inflationary evolution. A good choice might be the so-called Hubble slow-roll parameter set \cite{Liddle:1994dx}, which does not actually rely on the assumption of slow-roll in its definition. Instead, these parameters are defined entirely in terms of the Hubble scale $H(\phi)$ and its derivatives. These parameters reduce to the usual slow-roll expressions in the slow-roll limit. However, a generalization of these parameters is needed to include the DBI effect.
We shall introduce the corresponding inflationary parameters that are suitable here. We call them the DBI parameters. The analysis here follows closely that in Ref.\cite{Garriga:1999vw,Alishahiha:2004eh}.

We define expressions for the cosmological observables in terms of the Hubble parameter $H$, including factors of $\gamma$ to allow for relativistic-roll scenarios.
First we introduce the inflationary parameter $\epsilon_D$ (where the subscript  $D$ refers to DBI) as given by \cite{Alishahiha:2004eh}
\be
\label{0eps1}
\frac{\ddot{a}}{a}=H^2(1-\epsilon_D)
\ee
For inflation to occur $0<\epsilon_D<1$, so this is a good expansion parameter for observational quantities measured $N_{e}$ e-folds back from the end of inflation. By definition, inflation ends when 
$\epsilon_D=1$. Having defined $\epsilon_D$, the introduction of the other Hubble parameters follows naturally. We also need parameters to account for the variation of $\gamma$ with $\phi$. At first-order, we need the following three parameters

\ba
\epsilon_{D}&\equiv&\frac{2M_p^2}{\gamma}\left(\frac{H^{\prime}(\phi)}{H(\phi)}\right)^2\\\nonumber
\eta_{D}&\equiv&\frac{2M_p^2}{\gamma}\left(\frac{H^{\prime\prime}(\phi)}{H(\phi)}\right)\\\nonumber
\kappa_{D}&\equiv&\frac{2M_p^2}{\gamma}\left(\frac{H^{\prime}}{H}\frac{\gamma^{\prime}}{\gamma}\right)
\ea
We may define two additional parameters,
\ba
\xi_{D}&\equiv&\frac{4M_p^4}{\gamma^2}\left(\frac{H^{\prime}(\phi)H^{\prime\prime\prime}(\phi)}{H^2(\phi)}\right)\\\nonumber
\rho_{D}&\equiv&\frac{2M_p^2}{\gamma}\left(\frac{\gamma^{\prime\prime}}{\gamma}\right)
\ea
although we will see that these combinations only show up in derivatives of the first three. We will therefore consider only $\epsilon_D$, $\eta_{D}$, $\kappa_{D}$ and their derivatives.
We see that the presence of $1/\gamma$ in $\epsilon_D$ and the other parameters is the underlying reason why relativistic-roll inflation can generate enough e-folds.

Recall the usual slow-roll parameters:
\ba
\label{sr}
\eta_{SR}&\equiv& M_{p}^2\frac{V^{\pprime}}{V}  \\\nonumber
\epsilon_{SR}&\equiv&\frac{M_{p}^2}{2}\left(\frac{V^{\prime}}{V}\right)^2
\ea 
One may relate the above parameters for $\gamma \simeq 1$ to these slow-roll parameters.
In this case, let $T(\gamma -1) \simeq \dot{\phi}^{2}/2 = 2 M_{p}^{2}{H^{\prime}}^{2}$. We then have
\ba
\epsilon_{SR}=\epsilon_{D}\left(\frac{3-\eta_D}{3-\epsilon_D}\right)^2\\\nonumber
\eta_{SR}=\frac{1}{3-\epsilon_D}(3\epsilon_D + 3\eta_D + \eta_D^2-\xi_D)
\ea
So when $\epsilon_D \ll 1$ and $\eta_D \ll 1$, we have the slow-roll case, with 
$\epsilon_{SR} = \epsilon_D$ and $\eta_{SR} = \eta_D + \epsilon_D$. Slow-roll work frequently assumes that $\epsilon_{SR}\ll\eta_{SR}$, but it will be important for our discussion to keep in mind that this is not necessarily the case, and we must keep track of $\eta$, $\epsilon$, and $\kappa$.

The scalar density perturbation has been studied in 
Ref.\cite{Garriga:1999vw,Alishahiha:2004eh,Stewart:1993bc}.
Decomposing the inflaton into its rolling background $\phi(t)$ and a fluctuation $\delta$,
\ba
\phi= \phi(t) + \delta(x,t)
\ea
with a scalar perturbation (the Newtonian potential) $\Phi$ in the de-Sitter metric,
\be
ds^{2} = -(1 + 2 \Phi)dt^{2} +a(t)^{2}(1 - 2 \Phi)dx^{2}
\ee
it is easy to see that $\delta$ and $\Phi$ are related so that there is only one independent scalar fluctuation. The linear combination
$$\zeta = \frac{H}{\dot \phi} \delta + \Phi$$
becomes frozen as it exits the horizon during inflation, later generating the temperature fluctuation 
in the cosmic microwave background radiation. The evolution of $\zeta$ obeys a linearized Einstein equation. In terms of the variable 
\ba
\label{zdef}
z = \frac{a{\dot \phi} \gamma^{3/2}}{H}
\ea
one introduces the scalar density perturbation $u=\zeta z$. Introducing the conformal time $\tau$, 
$d\tau = dt/a$, we see that $u_{k}$ as a function of the wavenumber $k$ satisfies  
\ba
\label{ukeq}
\frac{d^{2}u_{k}}{d\tau^{2}} + \left( \frac{k^{2}}{\gamma^{2}} - \frac{1}{z} \frac{d^{2}z}{d\tau^{2}} \right) u_{k}=0
\ea
Note that the fluctuations $u$ travels at the sound speed $c_{s}$,
\ba
c_{s}^{2}= \frac{\partial p}{\partial \dot \phi} / \frac{\partial \rho}{\partial \dot \phi} = \frac{1}{\gamma^{2}}
\ea
so $u_{k}$ freezes when $k$ crosses $k=aH\gamma$ instead of $k=aH$. To solve for $u_{k}$, we express the potential in terms of inflationary properties. 
Following Eq(\ref{zdef}), we obtain
\ba
\label{d2zdt}
\frac{1}{z} \frac{d^{2}z}{d\tau^{2}} &=&  a^{2}H^{2} W  \\\nonumber
W &=&  2(1+\epsilon_{D}-\eta_{D}-\frac{\kappa_{D}}{2})
(1- \frac{\eta_{D}}{2} - \frac{\kappa_{D}}{4}) 
- \epsilon_{N} +\eta_{N} + \frac{\kappa_{N}}{2}
\ea 
where we introduce the derivative of the DBI parameters with respect to the e-fold number $N$, e.g.,
$\eta_{N}=\eta_{D},_{N}= \frac{d\eta_{D}}{dN}$,
\ba
\frac{\dot \phi}{H} \epsilon_{D}^{\prime} &=& - \epsilon_N = \epsilon_{D}( 2  \epsilon_{D} -2 \eta_{D} + \kappa_{D}) \\ \nonumber 
\frac{\dot \phi}{H} \eta_{D}^{\prime} &=& - \eta_{N}= \eta_{D}(\epsilon_{D} +  \kappa_{D}) - \xi_{D}  \\ \nonumber
\frac{\dot \phi}{H} \kappa_{D}^{\prime} &=& - \kappa_{N}= \kappa_{D} (2 \kappa_{D} +\epsilon_{D} - \eta_{D}) - \epsilon_{D}\rho_{D}
\ea
Expressing the Hubble scale in conformal time, $aH\tau(1-\epsilon_{D})= -1$,
Eq(\ref{ukeq}) becomes
\ba
\frac{d^{2}u_{k}}{d\tau^{2}} + \left( c_{s}^{2}k^{2} - \frac{\nu^{2}-1/4}{\tau^{2}} \right) u_{k}=0
\ea
where
$$\nu^{2} = \frac{W}{(1-\epsilon_{D})^{2}} + \frac{1}{4}$$
Here, $\nu \rightarrow 3/2$ as the DBI parameters vanish. Since all DBI parameters as well as $H$ vary much more slowly than $a(t)$, we may take $\nu$ to be close to constant, so the above equation behaves as a Bessel equation. We see that, for $aH \gamma \gg k$, the growing mode behaves as
\ba
|u_{k}| \rightarrow 2^{\nu - 2} \frac{\Gamma(\nu)}{\Gamma(\frac{3}{2})} \frac{1}{\sqrt{c_sk}} (c_sk\tau)^{1/2 -\nu} 
\ea
so the spectral density is given by
\ba
\label{spectralden} 
{\mathcal P}_R^{1/2}(k) = \sqrt{\frac{k^{3}}{2 \pi^{2}}} |\frac{u_{k}}{z}|
= 2^{\nu - 3/2} \frac{\Gamma(\nu)}{\Gamma(\frac{3}{2})} (1-\epsilon_{D})^{\nu -1/2} \frac{H^2}{2\pi |\dot{\phi}|}{\Large{|}}_{k=aH\gamma}
\ea
Then the scalar spectral index $n_{s}$ is given by $n_s-1=\frac{d\ln{\mathcal P}_R}{d\ln\:k}$ measured at $k=aH \gamma$. Since our expressions are functions of $\phi$, we rewrite
\ba
\frac{d}{d\ln\:k} = \frac{1}{\frac{d\ln\:k}{dt}} \dot \phi \frac{d}{d\phi} = \frac{1}{1 - \epsilon_{D} -\kappa_{D}}
\frac{\dot \phi}{H} \frac{d}{d\phi} =  \frac{-1}{1 - \epsilon_{D} -\kappa_{D}}\frac{d}{dN} 
\ea
Operating this on ${\mathcal P}_R$ (\ref{spectralden}), we obtain, keeping up to quadratic terms or equivalent,
\ba
\label{nsminus1}
n_{s}-1 &=& \frac{d\:\ln{\mathcal P}_R}{d\:\ln\:k} \\\nonumber
&=& (1-\epsilon_{D}-\kappa_{D})^{-1}(-4\epsilon_{D}+2\eta_{D}-2\kappa_{D} + 2 \epsilon_{N} +(4-2 \gamma_{E}-2 \ln 2)\nu_{N}) \\\nonumber
&\sim& (1+\epsilon_{D}+\kappa_{D})( -4\epsilon_{D}+2\eta_{D}-2\kappa_{D}) + 2 \epsilon_{N} -1.46 (2 \epsilon_{N} - \eta_{N} - \frac{\kappa_{N}}{2})
\ea 
Typically one needs only the first term in this expansion\footnote{We thank N. Agarwal for pointing out a typo in the last line of this equation.}. However, we will see that in the DBI case the first term exactly vanishes. It will be clear that any terms containing only derivatives ($\epsilon_N$, $\eta_N$, $\kappa_N$) will also vanish, and so in fact $n_s-1=0$ to all orders in this expansion in the ultra-relativistic approximation. 

The tensor mode spectral density, to first order, is given by
\ba
{\mathcal P}_h=\frac{2H^2}{M_p^2\pi^2}
\ea
and the corresponding tensor index:
\ba
n_t -1 &\equiv&\frac{d\ln\:{\mathcal P}_h}{d\ln\:k}\\\nonumber
&\approx&\frac{-2\epsilon_D}{1-\epsilon_D-\kappa_D}
\ea
This is non-vanishing even in the ultra-relativistic case. The ratio of power in tensor modes versus scalar modes is
\be
r=\frac{16\epsilon_D}{\gamma}
\ee
To keep $r \lesssim0.5$, we would like $\gamma$ to increase as $\epsilon_D$ does. However, the
 non-Gaussianity bound constrains $\gamma \lesssim 31$. This bound can be saturated under certain conditions in the intermediate regime. In the intermediate regime, it is possible to have $\epsilon_D\sim0.2$ and $\gamma\sim1$, so that $r$ exceeds the current bound.

The running of the spectral indices are given by
\ba
\frac{d\:n_s}{d\ln\:k}&=&\frac{-1}{(1-\epsilon_D-\kappa_D)^2}(-4\epsilon_N+2\eta_N-2\kappa_N+\dots)\\\nonumber
\frac{d\:n_t}{d\ln\:k}&=&\frac{-1}{(1-\epsilon_D-\kappa_D)^2}(-2\epsilon_N+\dots)
\ea
The scalar index running also exactly vanishes in the ultra-relativistic case.

\section{Slow-roll and ultra-relativistic limits}
We briefly review the key features of the slow-roll and DBI scenarios here, with the aim of embedding both possibilities in a single framework.

\subsection{Slow-roll inflation}

For small $m$, slow roll inflation is applicable. There are two regions of interest. First, let us consider the case with $m \ll H_{0}$. With small $\dot \phi$, the inflaton action simplifies so
\be
S= \int d^4x\:a^3(t)\left[ \frac{1}{2}\dot{\phi}^2 - V(\phi) \right]
\ee
With $\gamma \simeq 1$, we see that inflationary properties depend only on two parameters, namely $m$ and $V_{0}$ in $V(\phi)$. 
(Note that, for the model to be fully justified, $\phi_{i}$ should still be inside the throat. This imposes a relatively mild constraint on the remaining parameters.)
Recalling the usual slow-roll parameters (\ref{sr}),
we find that the final value of $\phi=\phi_{f}$ determined by $\eta_{SR} = -1$ is very close to $\phi_{A}$.
For all practical purposes, we shall simply set $\phi_{f}=\phi_{A}$. Alternatively, we could use the development of the tachyon (when the branes are within a string length) as the endpoint of inflation, but we find this makes very little difference. We can now find the initial value $\phi_{i}$ by going back 55 e-folds, calculating the density perturbation $\delta_{H}$ and fixing it to the COBE magnitude of
$\delta_{H}=1.9 \times 10^{-5}$. Other inflationary parameters such as the power index $n_{s}$, the tensor contribution $r$, the running of $n_{s}$ and the cosmic string tension $\mu$ are now functions of a single parameter $\beta = m^{2}/H_{0}^{2}$. They are given in Ref.\cite{Firouzjahi:2005dh}. We repeat here a few of the key equations for comparison to the DBI and general cases. These expressions assume that $V_0$ dominates the potential. First, there is a relationship between the slow-roll parameter $\eta_{SR}$ and $\beta$:
\ba
\eta_{SR}&=&\frac{\beta}{3}-\frac{5}{6N_e}\frac{1}{\Omega(\beta)}\\\nonumber
\Omega(\beta)&=&\frac{(1+2\beta)\e^{2\beta N_e}-(1+\beta/3)}{2\beta(N_e+5/6)(1+\beta/3)}
\ea
where $N_e$ is the number of e-folds. To fit the COBE normalization, we have
\ba
\delta_H&=&\left(\frac{2^{11}}{3\times5^6\times\pi^4}\right)^{1/6}N_e^{5/6}\left(\frac{T_3h_A^4}{M_p^4}\right)^{1/3}{\it f}(\beta)^{-2/3}\\\nonumber
{\it f}(\beta)&=&\left[\frac{2\beta(N_e+5/6)}{(1+2\beta)\e^{2\beta N_e}-(1+\beta/3)}\right]^{5/4}\frac{(1+2\beta)^{3/2}}{(1+\beta/3)^{1/4}}\e^{3\beta N_e}
\ea
The cosmic string tension is given by
\be
G\mu=\left(\frac{3\times5^6\times \pi^2}{2^{21}}\right)^{1/4}g_s^{-1/2}\delta_H^{3/2}N_e^{-5/4}f(\beta)
\ee
Finally, the scalar index is given by
\ba
\label{slowns}
n_s-1&\sim&2\eta_{SR}-6\epsilon_{SR}\\\nonumber
&\sim&\frac{2\beta}{3}-\frac{5}{3N_e}\frac{1}{\Omega(\beta)}-\frac{1}{3}\left(\beta+\frac{1}{2N_e\Omega}\right)^2(1+\beta/3)^{-1/3}\left(\frac{24N_e\phi_A^4\Omega}{N_AM_p^4}\right)^{1/3}
\ea
This is greater than zero for $\beta>0.01$ and increases with $\beta$. The slow roll approximations, ignoring $\epsilon_{SR}$, are valid up to about $\beta\sim1/7$, which is in the range of the bounds on the model from observation. Including $\epsilon_{SR}$, the slow-roll approximation is valid to about $m\sim10^{-6}M_p$.

For small $m$ with $m \gg H_{0}$, we find the usual chaotic inflation with a quadratic term. In this case,
$\delta_{H}$ fixes $m \simeq 6 \times 10^{-6}M_{p}$ and all the other inflationary parameters are determined. Note that this is the special case $\eta_{SR}=\epsilon_{SR}$.

\subsection{DBI inflation}

We consider DBI inflation in a single throat. If $m$ is large, so that the quadratic term dominates the potential, we can see from Eq.(\ref{potential}), Eq.(\ref{solve}) and Eq.(\ref{Happrox}) that we have the following approximate behavior in the large $\phi$ (late time) limit
\ba
\label{DBIapprox}
H(\phi)&\sim&\frac{\hat{m}}{M_p}\frac{\phi}{\sqrt{6}}\\\nonumber
\gamma(\phi)&\sim&2M_p^2\sqrt{\lambda}\frac{H^{\prime}(\phi)}{\phi^2}=\frac{\hat{m}M_p}{\phi^2}\sqrt{\frac{2\lambda}{3}}
\ea
where
\be
{\hat m}^{2} =m^2+2M_pm\sqrt{\frac{2}{3\lambda}}\left(1+\frac{2M_p^2}{3m^2\lambda}\right)^{1/2}+\frac{4M_p^2}{3\lambda}
\ee
Note that these definitions agree with the late-time behavior used in \cite{Silverstein:2003hf} and \cite{Alishahiha:2004eh}, up to a factor of $\sqrt{2}$ from the convention for $m$ and the replacement $m\rightarrow\hat{m}$. The expansion parameter $\epsilon_D$ is then given to lowest order by
\be
\hat{\epsilon}_{0}=\sqrt{\frac{6}{\lambda}}\frac{M_p}{\hat{m}}
\ee
where we use the hat to indicate that it is $\hat{m}$, not $m$ that enters the equation. The expression for the COBE normalization is
\be
\label{deltaHfr}
\delta_H\sim\frac{1}{30\pi}\frac{\hat{m}^2}{M_p^2}\sqrt{6\lambda}
\ee
These equations are of course approximate, and $\delta_H$ is not actually constant with $\phi$ when $\gamma$ is close to 1. However, for a given $m$, Eq.(\ref{deltaHfr}) determines the order of magnitude of $\lambda$.

With Eq.(\ref{DBIapprox}), we have, in the large $\gamma$ region,
\be
\eta_{D}, \xi_{D}, \epsilon_N, \eta_{N}, \kappa_{N} \rightarrow 0 \\\nonumber
\ee
(since $\kappa_{D}\rightarrow - 2\epsilon_{D},\rho_{D }\rightarrow 6\epsilon_{D}$ ) 
so that Eq.(\ref{d2zdt}) becomes
\ba
\frac{1}{z} \frac{d^{2}z}{d\tau^{2}} \rightarrow  a^{2}H^{2}(2 + 5  
\epsilon_{D} +  2\epsilon_{D}^{2})
\ea
In fact, if we keep higher order terms, we see that only $\epsilon_N$, $\eta_{N}$, and $\kappa_{N}$ come in, so that $n_s-1 \rightarrow 0$.

We now find the first correction to Eq.(\ref{DBIapprox}). When the $V_0$ term can be neglected, we have (from Eq.(\ref{solve}))
\ba
3M^{2}H^{2} &=& \hf m^{2}\phi^{2} + T(\gamma-1)\\\nonumber
\gamma &=& \frac{2M^{2}H'}{\sqrt{T}}(1+\dots)
\ea
and recall $\lambda T= \phi^{4}$.
We can write
\ba
\label{expansion}
H(\phi)&=&A\phi(1 - B \phi^{2}+C\phi^4+\dots)\\\nonumber
\gamma(\phi) &=& a(\frac{1}{\phi^{2}} - b+c\phi^2+\dots)
\ea
Ignoring all terms except those containg just $a$ or $A$, we recover Eq.(\ref{DBIapprox}). For small $\phi$, we may keep $b$ and $B$ consistently and solve for all four coefficients.
\ba
A&=&\frac{\hat m}{\sqrt{6} M_p}\\\nonumber
a&=&2A M_p^{2} \sqrt{\lambda}=\hat{m}M_p\sqrt{\frac{2\lambda}{3}}=M_pm\sqrt{\frac{2\lambda}{3}\left(1+\frac{2M_p}{3\lambda m^2}\right)}+\frac{2M_p^2}{3}\\\nonumber
b&=&3B\\\nonumber
B&=& \frac{1}{\lambda {\hat m}^{2}-3a}=\frac{1}{\lambda\hat{m}^2-\hat{m}M_p\sqrt{6\lambda}}
\ea
We find that the first correction to $n_s-1$ in the large gamma and small $\phi$ limit is positive and is given by
\be
n_s-1=\frac{8\hat{\epsilon}_0^2}{3\hat{\gamma}_0(1-\hat{\epsilon}_0)}(1+1.42\hat{\epsilon}_0)
\ee
with 
\be
\hat{\gamma}_0=\frac{a}{\phi^2}=\frac{\hat{m}M_p}{\phi^2}\sqrt{\frac{2\lambda}{3}}
\ee
This expression is good only for large gamma and small $\phi$, and for the particular choice of $h(\phi)$ in Eq.(\ref{warpfunction}). We frequently find the 55 e-fold condition at large $\phi$ and relatively small $\gamma$, so that $n_s-1$ is negative rather than postive. However, the above correction may be seen at fewer e-folds, once $\gamma$ has become significantly larger than 1. One may also work out the $\phi$ dependence in the number of e-folds and the expression for the COBE normalization. Note that   Ref.\cite{Alishahiha:2004eh} is concerned with obtaining of order 20 e-folds, which is certainly in the large $\gamma$ regime. The expanision in Eq.(\ref{expansion}) is useful for solving the differential equations for $\gamma(\phi)$ and $H(\phi)$ because it gives boudary conditions valid at the end of inflation. For a more general complicated function $h(\phi)$, it may be easier to instead use the behavior of Eq.(\ref{solve}) at very large $\phi$, where $\gamma\rightarrow1$ and the $\gamma$-dependent corrections to $H(\phi)$ vanish.

\section{Between slow-roll and DBI}

As mentioned earlier, the value of $m$ is sensitive to the details of the particular string  vacuum one is considering and to the various stringy/quantum corrections that are present. In all of these cases, determining the precise potential, complete with corrections, is a difficult task. Many issues and their possible relevance for cosmology have been discussed (see for example 
Ref.\cite{McAllister:2005mq,Shandera:2004zy}). Here we roll all of this complicated physics into the $\frac{1}{2}m^2\phi^2$ term. We could expand this analysis to include terms that might enter in different scenarios.

We have discussed the cases where $m\ll H_0$ (slow-roll) and where the quadratic term dominates the potential (DBI). The intermediate mass range is difficult to treat analytically. Here we discuss the qualitative features and give numerical results. Note that to ensure inflation in the intermediate region, it so that 
\ba
\label{Happrox}
H^2&\sim&\frac{1}{3M_p^2}V\\\nonumber
H^{\prime}&\sim&\frac{1}{2\sqrt{3}M_p}\frac{V^{\prime}}{V^{3/2}}
\ea

\subsection{Understanding the Parameter Space}
We have four parameters (fixing $g_s$) to work with. We choose the first two, $m$ and $V_0$, by  writing the potential in a simple form
\be
V(\phi)=\hf m^2\phi^2+V_0\left[1-\frac{cV_0}{\phi^4}\right]
\ee
where $c=27/(64\pi^2)$ and $V_0=2T_3h_A^4$. The other two parameters are related to the basic requirement of inflation - sufficient e-folds. To calculate $N_e$, we use
\be
\label{e-folds}
N_e=\int H\:dt=\int_{\phi_i}^{\phi_f}\frac{H}{\dot{\phi}}\:d\phi
\ee
The endpoint of inflation comes shortly before the brane and anti-brane annihilate, when the tachyon develops. However, we can set $\phi_f=\phi_A=\lambda^{1/4}h_A$ for simplicity. We also require inflation to take place entirely within the throat, so that $\phi_i<\phi_{edge}=\sqrt{T_3}\:R=(T_3\lambda)^{1/4}$. Realistically, some inflation may come from other parts of the $D$3's trajectory, but the potential and the metric must be modified to treat those regimes. The quadratic piece in the potential contains the most complicated physics, but we see that, details aside and for fixed $\ap$ and $g_s$, $m$ controls which regime we are in: for large $m$ ($m>10^{-5}M_p$), the quadratic piece dominates and we have DBI inflation. This looks a bit like chaotic inflation, but notice that the DBI action led to factors of $\gamma$ in the equation of motion. That change makes this regime much more flexible and interesting. For $m$ small ($m<10^{-7}M_p$) we have either the usual slow-roll inflation, where the constant term dominates, or chaotic inflation for small enough $V_0$. For intermediate $m$, the constant piece generically dominates, but the quadratic and Coulomb terms also play an important role. In this regime it is more difficult to make an analytic approximation. In terms of parameters, the slow-roll regime is controlled by $\eta_D$. Toward the intermediate regime, $\epsilon_D$ grows and with $\eta_D$ determines the behavior of the observables. The parameter $\kappa_D$ is important in the relativistic and ultra-relativistic regimes.

\subsection{Predictions for Observables}

Imposing the magnitude of the density perturbation as determined by COBE, 
$\delta_{H} = \delta \rho/\rho \simeq 1.9 \times 10^{-5}$, we find that naively $m_{0} \simeq   10^{-7}M_p $.
We plot values of $m$ around that value. Note that we really have a four-dimensional parameter space (fixing $g_s$). We use the COBE data to effectively fix one additional parameter, but this means that along our two dimensional plots, the values of $\lambda$, $h_A$ and $m_s$ can vary. In general we fix $m_s$ along a line unless otherwise noted. Let us summarize the key results of the analysis here.

\begin{itemize}

\item {\bf Scalar Spectral Index:} 

The most important parameter in an inflationary scenario is probably the scalar power index $n_{s}$. The value $n_{s}-1$ measures the deviation of the density perturbation from a scale-invariant power spectrum. We give the values of $n_s-1$ as a function of $m/M_p$ in Fig.(\ref{nsplot}). To understand the behavior of this plot, it is usful to keep in mind the first three terms from Eq.(\ref{nsminus1}), which gives $n_s-1\sim2\eta_D-4\epsilon_D-2\kappa_D$. Keeping $m_s$ and $g_s$ fixed, increasing $m/M_p$ corresponds to increasing $\gamma$. For very small $m$, $\eta_{D}<0$ and $n_{s} \simeq 0.97$ \cite{Kachru:2003sx}. As $m$ increases in the slow roll region, $n_{s}$ increases to larger than 1 (i.e, blue tilt, $\eta_{D}>0$ ). The dashed line shows the analytic approximation in the slow-roll region, Eq.(\ref{slowns}), assuming $V_0$ dominates the potential. The actual values (points) drop below this line as the quadratic term becomes important. The trend toward a red tilt can be understood as similar to chaotic inflation as the size of the quadratic term grows. In chaotic inflation the slow roll parameters $\epsilon_{SR}$ and $\eta_{SR}$ are equal, so that $n_s-1\sim2\eta_{SR}-4\eta_{SR}$ is negative. Around $m/M_p\sim3\times10^{-6}$ the slow-roll approximation breaks down. The parameter $\kappa_D$ grows and the spectal index turns back toward zero. Even though slow-roll may be valid at $N_{e}=55$ e-folds before the end of inflation, the inflaton can become relativistic as it falls down the throat so a treatment beyond the usual slow-roll approximation is needed. In particular, the number of e-folds depends quite sensitively on this. In principle the curve matches smoothly onto the ultra-relativistic case (where $n_s-1=0$). It is difficult to match the COBE normalization at large $m/M_p$, but we can instead see this behavior by adjusting $h_A$ and $\lambda$, and $\ap$ for a particular mass to obtain large $\gamma$. The arrowed line shows that the spectral index returns to 1 as $\gamma$ increases. We show in \S4.4 the behavior of the parameters $\eta_D$, $\epsilon_D$, and $\kappa_D$ at 55 e-folds as a function of $m$.

\begin{figure}[htb]
\begin{center}
\includegraphics[width=0.7\textwidth,angle=0]{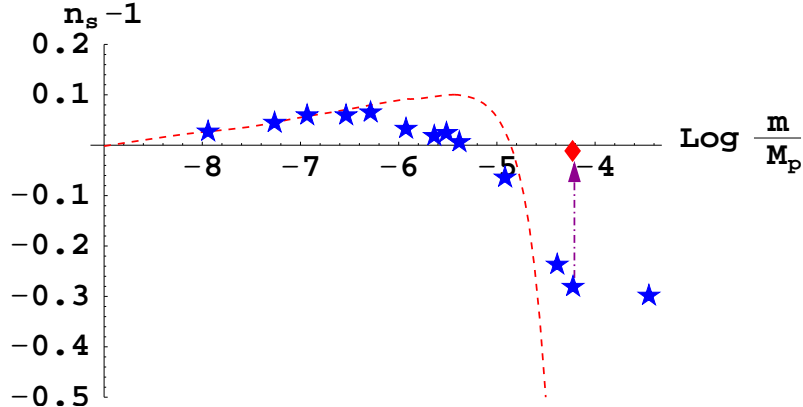} 
\caption{The scalar index as a function of inflaton mass. The dashed line is the slow-roll analytic prediction, assuming $V_0$ dominates the potential. The arrowed line demonstrates that $n_s-1\rightarrow0$ as $\gamma$ increases. The diamond point is for $\gamma=22$, near the current bound from non-Gaussianities.
\label{nsplot}}
\end{center}
\end{figure}

\item {\bf Scalar Tensor Ratio and Running:}
Various other observables take interesting values in certain regions of the parameter space. The running of $n_{s}$ and the ratio of the tensor to the scalar mode $r$ are shown in Fig.(\ref{dnsplot}) and Fig.(\ref{rplot}). The slow-roll behavior for the running of the spectral index, assuming $V_0$ dominates the potential, is shown by dashed lines. Note that $dn_s/d\ln k\rightarrow0$ in the ultra-relativistic case. The tensor/scalar ratio $r$ can be quite large in the intermediate and ultra-relativistic regions. This is because $\epsilon_D$ grows as the quadratic term starts to dominate over the constant term in the potential. Growth in $\gamma$ eventually pushes $r$ small again. WMAP finds $r<0.5$.  

\begin{figure}[htb]
\begin{center}
\includegraphics[width=0.7\textwidth,angle=0]{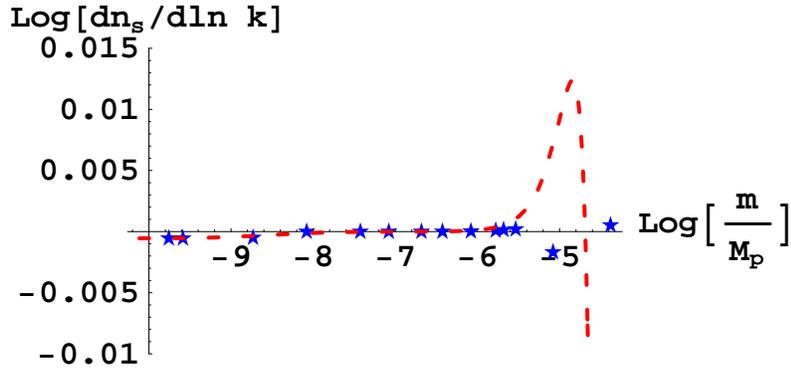} 
\caption{The running of the scalar index as a function of inflaton mass. The dashed line is the first order slow-roll result, assuming $V_0$ dominates the potential.
\label{dnsplot}}
\end{center}
\end{figure}

\begin{figure}[htb]
\begin{center}
\includegraphics[width=0.7\textwidth,angle=0]{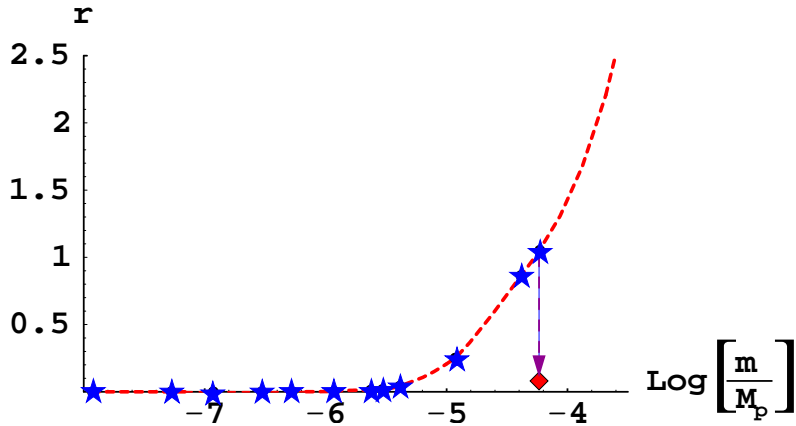} 
\caption{The ratio of power in tensor modes to power in scalar modes. The arrowed line shows the decrease in $r$ to near zero as $\gamma$ increases. The point at the bottom of the arrow has $\gamma\sim20$. 
\label{rplot}}
\end{center}
\end{figure}

\item {\bf Non-Gaussianity:}
In the slow-roll case, we expect non-Gaussianity to be negligible, but for large $\gamma$ it is shown in Ref \cite{Alishahiha:2004eh} that
\be
{\it f}_{NL}\approx(0.32)\gamma^2
\ee
The current bound on ${\it f}_{NL}$ is weak, but constrains $\gamma$ to be less than about 31. Ref.\cite{Creminelli:2005hu} has also examined WMAP constraints for DBI-type models in detail. Large $\gamma$ generally comes with increasing $\lambda$ and decreasing $h_A$ in the intermediate regime. It usually corresponds to $n_s-1<0$, 

\end{itemize}

\subsection{Number of e-folds}

The basic criterion for inflation is that we must obtain enough e-folds. The number of e-folds is defined in Eq.(\ref{e-folds}). Using Eqs.(\ref{phidot}) and (\ref{solve}), we can write
\ba
N_e&=& -\frac{1}{2M_p^2}\int_{\phi_i}^{\phi_f}\frac{H(\phi)\gamma(\phi)}{H^{\prime}(\phi)}\:d\phi\\\nonumber
&=&-\frac{1}{M_p^2}\int_{\phi_i}^{\phi_f}\gamma(\phi)\left[\frac{V}{V^{\prime}}-\frac{T_3h^4(1-\gamma(\phi))}{V^{\prime}}\right]\left[1-\frac{\eta_{D}}{3}\right]d\phi\\\nonumber
&\sim&-\frac{1}{M_p^2}\int_{\phi_f}^{\phi_i}\gamma(\phi)\frac{V}{V^{\prime}}\:d\phi\\
\ea
From here we can see that the relativistic case essentially benefits by a factor of $\gamma$ in the number of e-folds. As we saw in previous sections, $\gamma(\phi)$ becomes large rapidly as the brane nears the end of the throat, so the acutal number of e-folds can be quite large.

\begin{figure}[htb]
\begin{center}
\includegraphics[width=0.7\textwidth,angle=0]{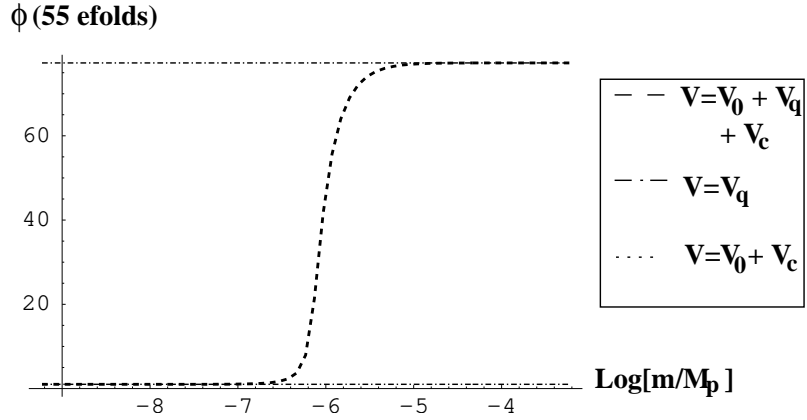} 
\caption{The value of $\phi$ (units of $\alpha^{\prime-1/2}$) that gives 55 e-folds as a function of $m/M_p$. The dashed lines shows the result for the full potential, while the upper and lower dot-dashed lines show potentials with quadratic and constant + Coulombic terms respectively.
\label{Nefoldsplot}}
\end{center}
\end{figure}

As a general rule, DBI inflation requires a longer throat to obtain enough e-folds. In Fig. \ref{Nefoldsplot} we show how the length of throat needed to obtain 55 e-folds ($\phi_{55}$) varies with $m/M_p$. The plot is for a particular choice of $\lambda$ and $V_0$, so it does not incorporate the changes in those parameters necessary to match the COBE data. However, one can easily see the general feature that much of the mass range is like either the slow-roll case (lower dot-dashed line shows a potential without the quadratic piece) or the DBI case (upper dot-dashed line shows an $m^2\phi^2$ potential). There is a small interpolating region between (dashed line shows the full potential). 
Of course, as discussed earlier, 55 e-folds before the end of inflation may occur while the brane is moving out of a throat or is in the bulk. These cases may require only a simple generalization of what we have done here. For example, a quartic term may be added to the potential.

If there were a scenario where we were interested in obtaining just the first 20 or so e-folds from a model of this type, our predictions would be very different due to the sharp increase in $\gamma$. Such an example seems quite contrived, but it is interesting to note that in that case we could find much larger non-Gaussianities in a larger region of parameter space.

\subsection{Constraints from $\delta_H$}

The shape of the $\delta_H$ curve changes significantly with the relative sizes of terms in the potential. It is interesting to see how the COBE condition comes in in the different regimes. In terms of $H(\phi)$, we have
\be
\label{deltahH}
\delta_H=\frac{H(\phi_i)}{30\pi M_p^2H^{\prime}(\phi_i)}(3\gamma H(\phi_i)-2M_p^2H^{\prime\prime}(\phi_i))
\ee 
This expression comes from starting with 
\be
\delta_H=C\frac{\delta\phi V^{\prime}}{\rho+p}
\ee
The constant $C$ is determined by matching the $\gamma\sim1$ case to the slow-roll expression
\be
\label{deltahsr}
\delta_H=\frac{1}{\sqrt{75}\pi M_p^3}\frac{V^{3/2}}{V^{\prime}}
\ee
using Eq.(\ref{Happrox}). 

The approximate expressions for the slow-roll and DBI cases show that the two cases do not depend on the same parameters with the same sensitivity. In the DBI regime (Eq.(\ref{deltaHfr})), $\lambda$ is roughly fixed, while $V_0$ may vary over an order of magnitude. In the slow-roll case, $\delta_H$ depends on $V_0$, but $\lambda$ is only bounded by the requirement that we stay in the throat.

When the quadratic term in the potential grows very large, we might expect to easily find the ultrarelativistic case with very large $\gamma$ possible. We have Eq.(\ref{deltaHfr}) valid for small $\phi$, large $\gamma$, which we can in principle solve for a value of $\lambda$ such that the COBE normalization is satisfied. But for large enough $m$, $\hat{m}$ is so large that no solution is possible. The further $\phi$-dependent corrections are not quite enough to compensate unless one moves the edge of the throat out by decreasing $\ap$ or $g_s$. But moving out in $\phi$ means moving to smaller $\gamma$, so that in the end we still find $\gamma\sim2$ at the most. 

\subsection{Numerics}
\begin{figure}[htb]
\begin{center}
\includegraphics[width=0.7\textwidth,angle=0]{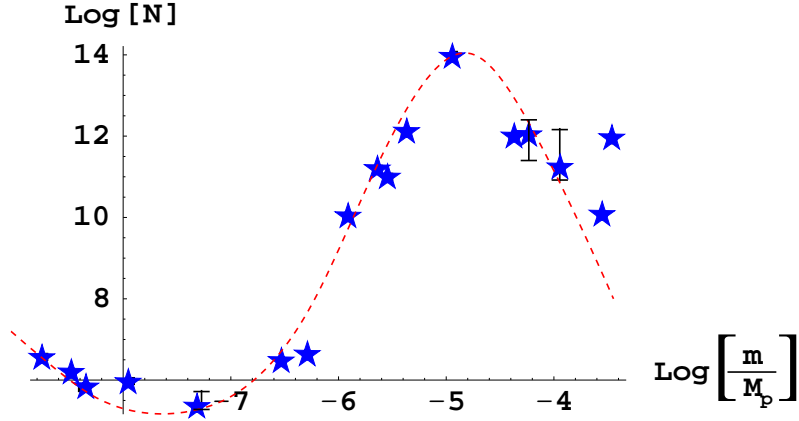} 
\caption{Range of values of the $D$3 charge (N) that give working inflation models. The dashed line shows the trend of the points, and the bars show the range $N$.
\label{mNplot}
}
\end{center}
\end{figure}

In the intermediate regime, there are no simple analytical approximations to relate the parameters. We show in Table \ref{paramrange} the range of values for $V_0$ and $N$ in the three mass ranges. Keep in mind that there is not necessarily a linear correspondence along this range. Fig.(\ref{mNplot}), for example, shows the values of $N$ that work along the mass range (but $V_0$ is also changing).

The range of cosmic string tensions shown in Fig.(\ref{Gmuplot}) comes from the variation possible in $N$ ($\lambda$). 55 e-folds of inflation can be obtained for a range of $N$ if we also vary $V_0$. Fig. (\ref{mNplot}) shows the behavior of $N$ with $m$. We include some sample points for reference.

\begin{table}
\caption{\label{paramrange}Range of parameters that give enough e-folds of inflation and match the COBE data. We have fixed $g_s=0.1$ and $m_s=10^{-2}M_p$.}
\begin{center}
\begin{tabular}{@{}|c|c|c|c|}
\hline
 & {\bf m}& {\bf $V_0$} & {\bf N} \\
\hline
{\bf Slow-roll} & $m<10^{-7}M_p$ & $10^{-12}\lesssim V_0\lesssim10^{-8}$ & $10^5\lesssim N\lesssim10^7$\\
\hline
{\bf Intermediate} & $5\times10^{-6}M_p<m<10^{-7}M_p$ & $10^{-12}\lesssim V_0\lesssim 10^{-8}$& $10^{6}\lesssim N\lesssim 10^{12}$\\
\hline
{\bf DBI} & $m>5\times10^{-6}M_p$ & $10^{-25}\lesssim V_0\lesssim10^{-12}$ & $10^{9}\lesssim N\lesssim10^{14}$\\
\hline
\end{tabular}
\end{center}
\end{table}

\begin{table}
\caption{\label{points}Sample points with the corresponding plot symbols used in Figures 1, 2, and 3. The third and eighth points are ruled out by their large $r$ values. The first two points have $\gamma$ at 55 e-folds equal to 1 to six decimal places. The value of $\gamma$ at one e-fold is 1.00003 for point 2 and 80,000 for point 5.}
\begin{center}
\begin{tabular}{@{}|c|c|c|c|c|c|c|c|c|c|c|}
\hline
{\bf Point}&${\bf m/M_p}$ & ${\bf N}$&${\bf hA}$&${\bf \ap M_p^2}$&${\bf V_0}$&
${\bf \gamma_{55}}$&${\bf n_s}$&${\bf r}$&${\bf \frac{dn_s}{d\ln k}}$&${\bf G\mu}$ \\
\hline
{\bf 1}($\bigstar$)&$1\cdot10^{-8}$& $9\cdot10^5$& $0.024$& $1600$& $10^{-13}$& $1$& $1.025$& $10^{-7}$& $2\cdot10^{-9}$&$10^{-8}$\\
\hline
{\bf 2}($\bigstar$)&$1\cdot10^{-6}$& $1\cdot10^{10}$& $0.227$& $1600$& $10^{-9}$& $1$& $1.031$& $10^{-5}$& $2\cdot10^{-5}$&$10^{-6}$\\
\hline
{\bf 3}($\bigstar$)&$6\cdot10^{-5}$& $1\cdot10^{12}$& $0.0002$& $1600$& $10^{-21}$& $1.22$& $0.72$& $1.05$& $5\cdot10^{-4}$& $10^{-12}$\\
\hline
{\bf 4}($\blacklozenge$)&$6\cdot10^{-5}$& $2.57\cdot10^{12}$& $0.0077$& $10^4$& $10^{-16}$& $2.61$& $0.939$& $0.50$& $1\cdot10^{-2}$& $10^{-10}$\\
\hline
{\bf 5}($\blacklozenge$)&$6\cdot10^{-5}$& $3.02\cdot10^{12}$& $0.0050$& $10^4$& $10^{-17}$& $11.7$& $0.996$& $0.11$& $7\cdot10^{-4}$& $10^{-10}$\\
\hline
{\bf 6}($\blacklozenge$)&$6\cdot10^{-5}$& $3.03\cdot10^{12}$& $0.0049$& $10^4$& $10^{-17}$& $13.94$& $0.997$& $0.093$& $5\cdot10^{-4}$& $10^{-10}$\\
\hline
{\bf 7}($\diamondsuit$)&$6\cdot10^{-5}$& $3.07\cdot10^{12}$& $0.0024$& $10^4$& $10^{-18}$& $48.5$& $0.999$& $0.027$& $8\cdot10^{-5}$& $10^{-11}$\\
\hline
{\bf 8}($\triangle$)&$1\cdot10^{-4}$& $1\cdot10^{11}$& $3\cdot10^{-5}$& $1600$& $10^{-24}$& $3.096$& $0.918$& $0.85$& $4\cdot10^{-2}$& $10^{-14}$\\
\hline
\end{tabular}
\end{center}
\end{table}

Some of our final numbers differ from \cite{Alishahiha:2004eh}. This is due to the more precise constraints we place on the initial and final values of $\phi$ and the number of e-folds obtained. Since $\gamma$ decreases quickly with increasing $\phi$, requiring more e-folds decreases the value of $\gamma$ at the beginning of inflation. For the purposes of matching non-Gaussianity bounds, we need to be sure $\gamma$ stays less than about 30 for the 10 or so e-folds we see. Fig.(10) shows how $\gamma$ and the total number of e-folds correspond. It also shows the growth of $\gamma$ as the brane moves down the throat.

\begin{figure}[h]
\begin{center}
$\begin{array}{c@{\hspace{1cm}}c}

\epsfxsize=8cm
\epsffile{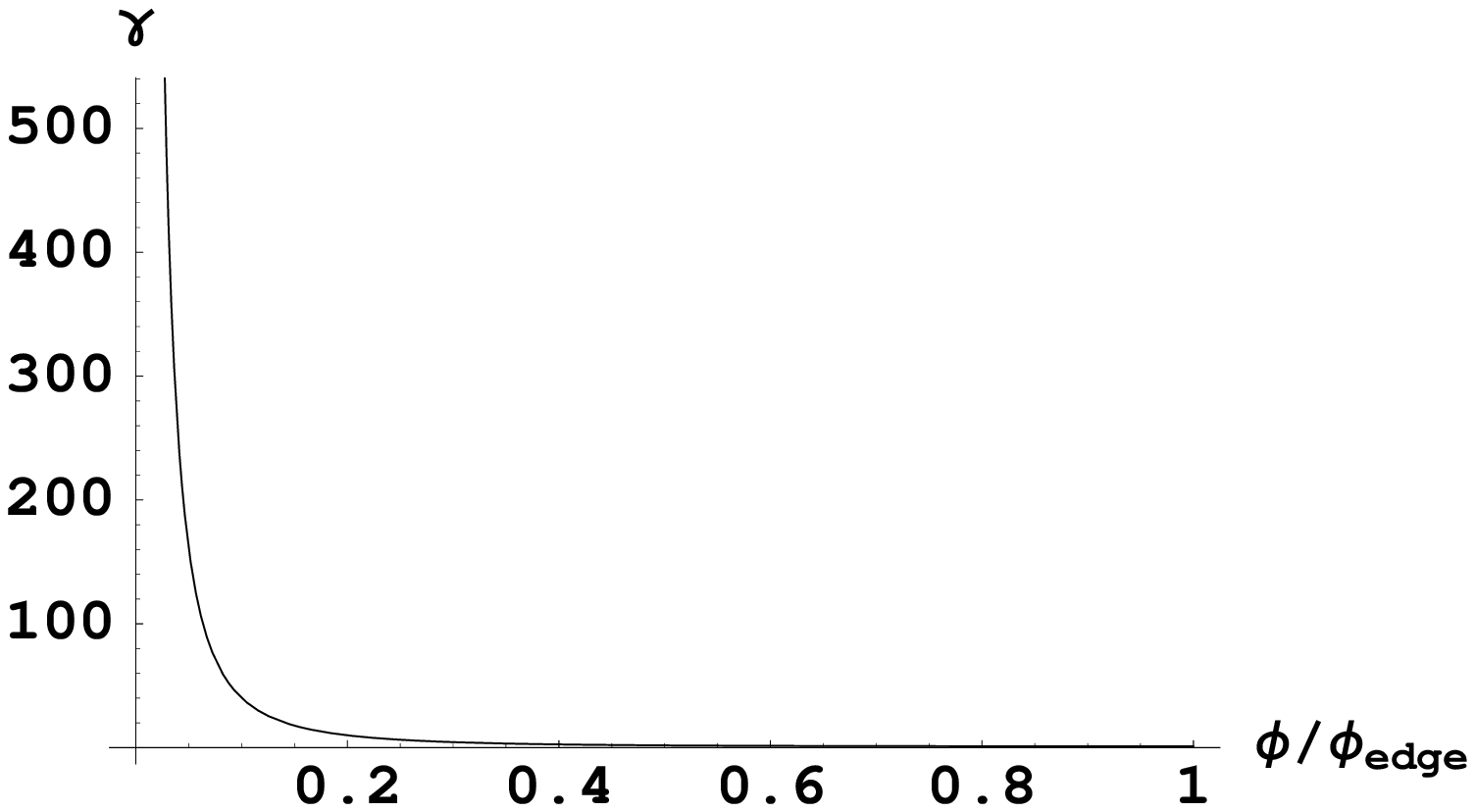} &
	\epsfxsize=7cm
	\epsffile{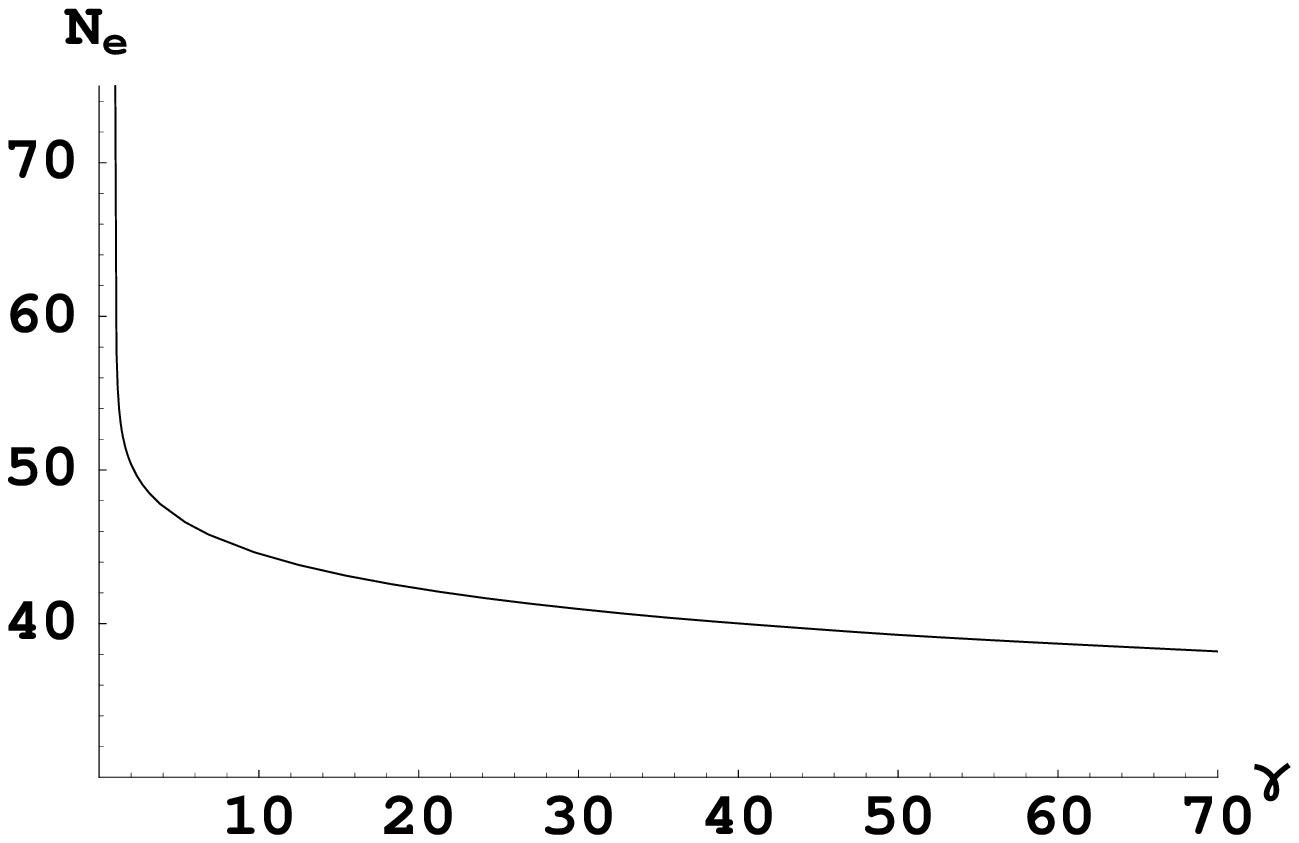} \\ [0.4cm]
\mbox{\bf (a)} & \mbox{\bf (b)}
\end{array}$
\end{center}
\caption{These plots illustrate the growth in $\gamma$ as inflation progresses. Plot (a) shows the sharp increase of $\gamma$ as the brane moves toward the bottom of the throat. (b) replots this information as $\gamma$ as a function of e-fold number.}
\label{doubleplot1}
\end{figure}

\begin{figure}[htb]
\begin{center}
\includegraphics[width=0.6\textwidth,angle=0]{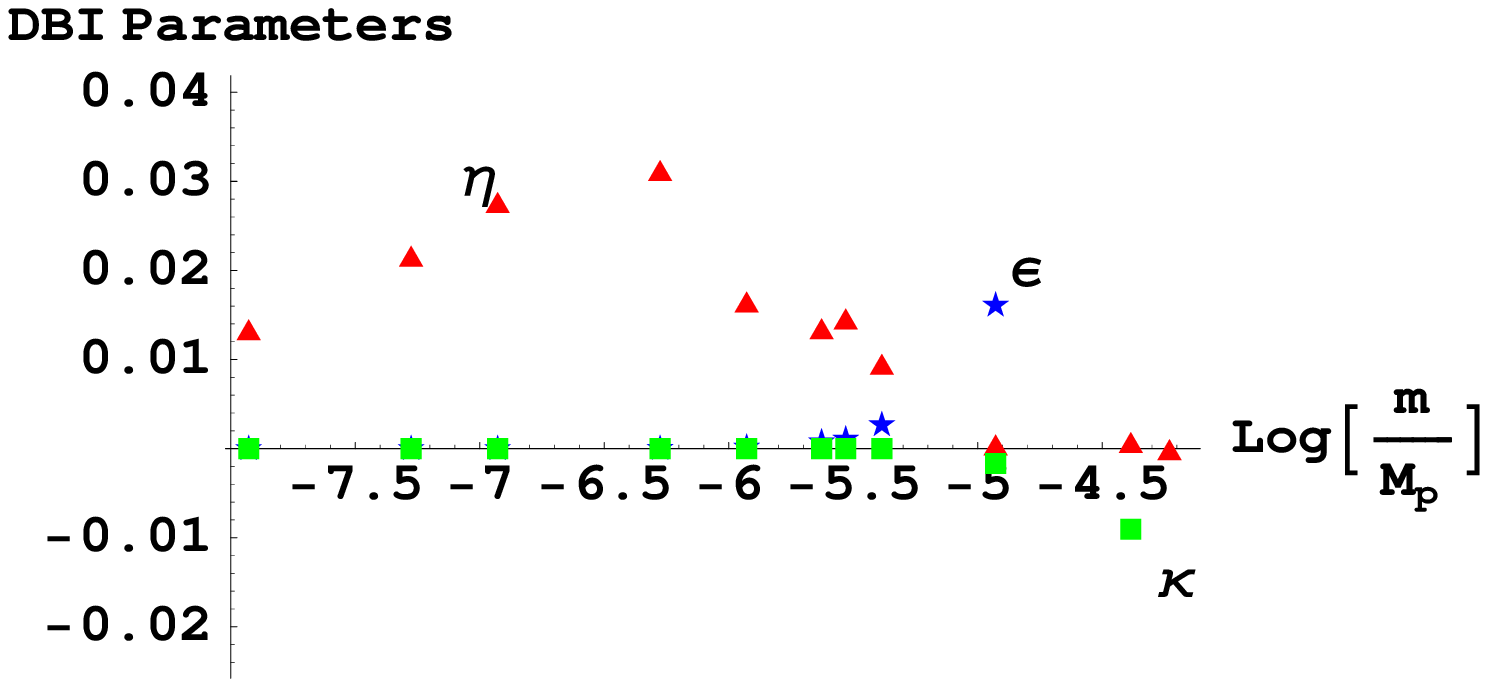} 
\caption{The value of $\epsilon_D$ (stars), $\eta_D$ (triangles), and $\kappa_D$ (boxes) at 55-efolds, for $\gamma\sim1$.
\label{paramsm}}
\end{center}
\end{figure}

\begin{figure}[htb]
\begin{center}
\includegraphics[width=0.6\textwidth,angle=0]{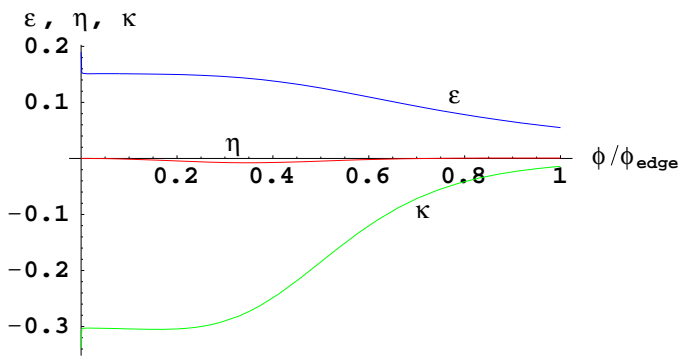} 
\caption{The behavior of the parameters with $\phi$ for $m=5.89\times10^{-5}M_p$. Here $\gamma\sim1$ at 55 e-folds, but grows sharply near the end of inflation.
\label{paramsphi}}
\end{center}
\end{figure}

We can gain some insight into the behavior of the observables in the intermediate region by examining the behavior of the parameters $\epsilon_D$, $\eta_D$, and $\kappa_D$. Fig.(\ref{paramsm}) and Fig.(\ref{paramsphi}) show the values of these parameters at 55 e-folds for several values of $m$ and the $\phi$-evolution of the parameters for a particular case in the intermediate regime. There $\gamma$ is very nearly one for roughly the upper third of the throat and $\epsilon_D$ dominates while $\eta_D$ and $\kappa_D$ are very small. As $\gamma$ grows (see Fig.10), $\kappa_D$ becomes more important. From the plot, we see that the magnitude of $\kappa_D$ increases to twice the value of $\epsilon_D$, as expected for large $\gamma$.

For the same value of $m$, the first plot in Fig.(12) shows the number of e-folds as a function of position in the throat. The second plot shows that, for all parameters fixed except $h_A$, the number of e-folds at the $\phi$ value where the correct COBE normalization is found changes linearly with $h_A$, although the $\phi$ value does not change. This is convenient, since the number of e-folds corresponding to the scale of the COBE normalization (the horizon size) depends on the reheating temperature. Once this temperature is calculated in string theory models, we should be able to match precisely by varying $h_A$.

\begin{figure}[h]
\begin{center}
$\begin{array}{c@{\hspace{1cm}}c}
\epsfysize=5cm
\epsffile{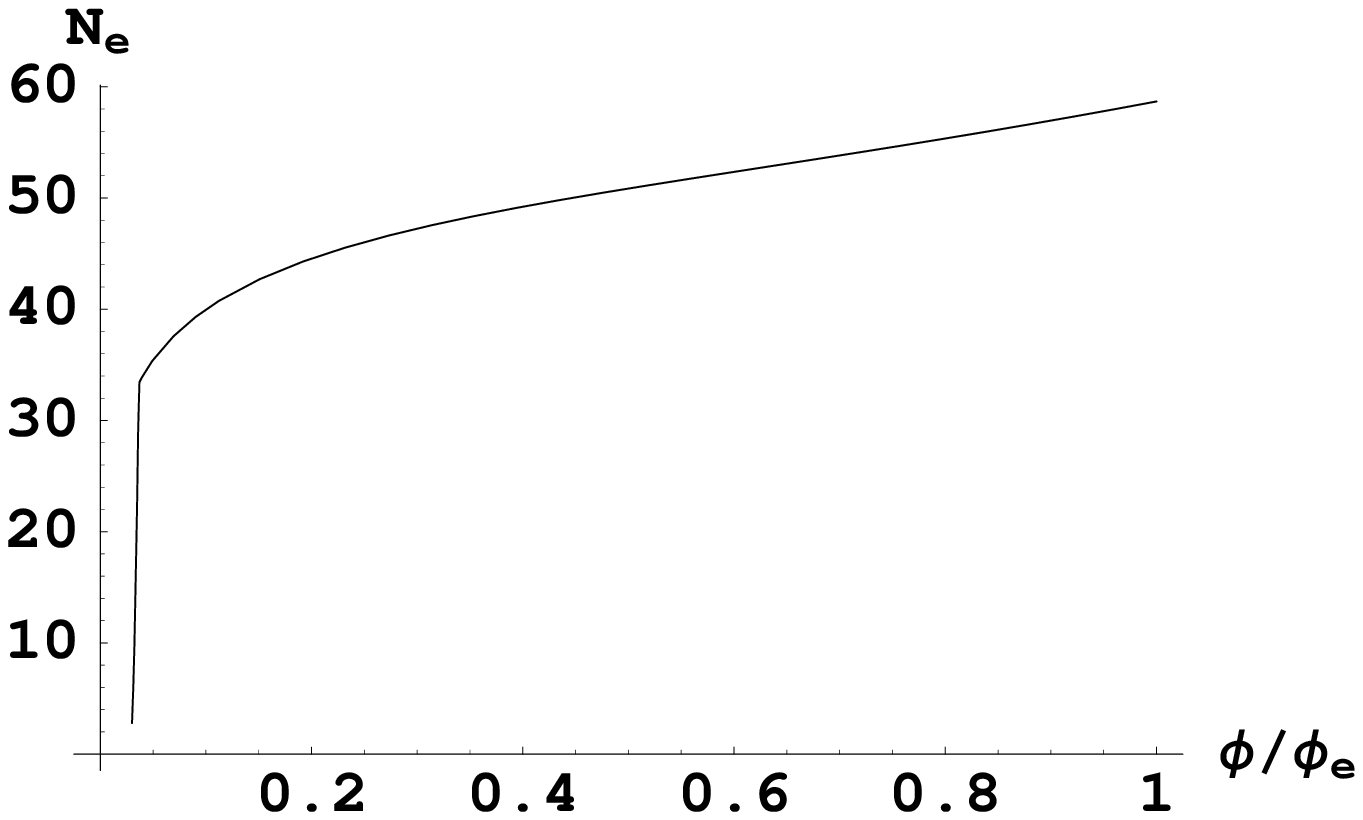} &
	\epsfysize=5cm
	\epsffile{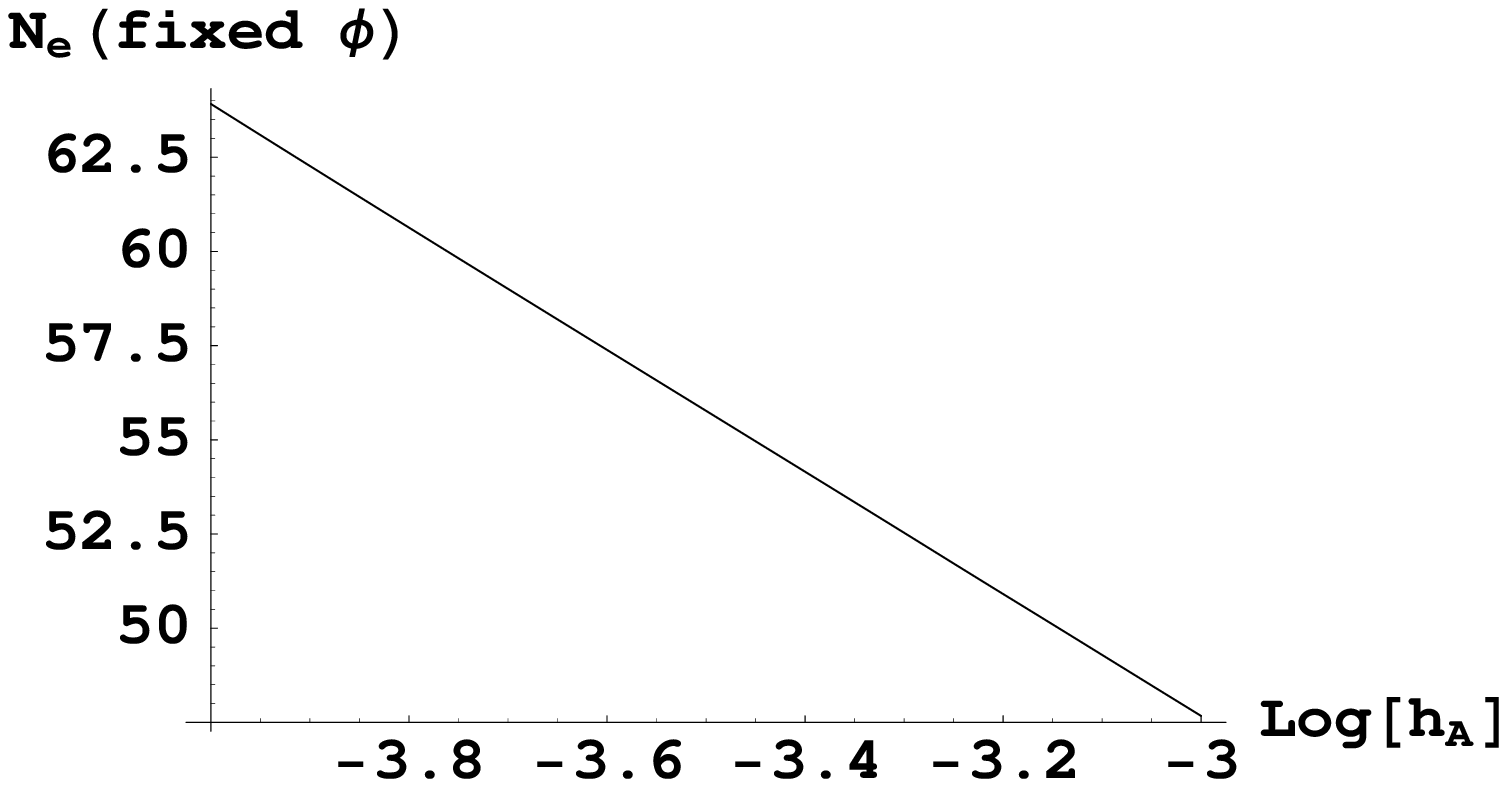} \\ [0.4cm]
\mbox{\bf (a)} & \mbox{\bf (b)}
\end{array}$
\end{center}
\caption{Plot (a) shows the number of e-folds as a function of position in the throat. Plot (b) shows the shift in number of e-folds corresponding to the fixed $\phi$ position where $\delta_H=1.9\times10^{-5}$ is obtained as $h_A$ is varied.}
\label{doubleplot3}
\end{figure}

The cosmic string tension $\mu$ is given by ($G^{-1}=8 \pi M_{p}^{2}$)
\be
G\mu=GT_1h_A^2=\sqrt{\frac{1}{32\pi g_s}}\left(\frac{T_3h_A^4}{M_p^4}\right)^{\hf}
\ee
The cosmic string tension increases with $m$ for the slow-roll case, but then decreases in the relativistic regime, where $T_3h_A^4$ must be much smaller to give enough e-folds (Fig. (\ref{Gmuplot})). The bars in the figures indicate the range of predicted values due to the range of allowed warp factor.

\begin{figure}[htb]
\begin{center}
\includegraphics[width=0.7\textwidth,angle=0]{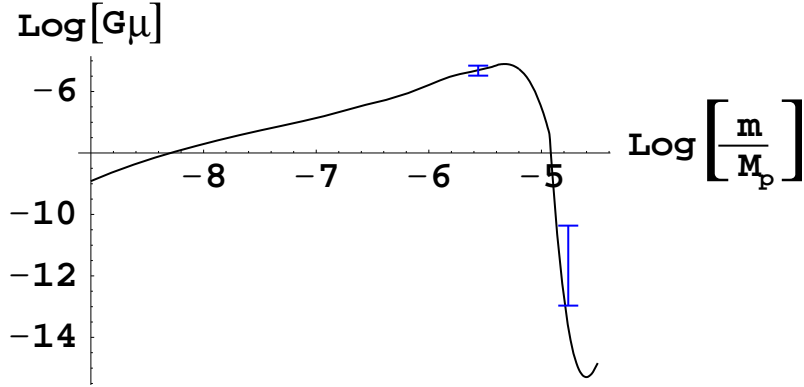} 
\caption{Cosmic string tensions predicted (for fixed string scale $m_s\sim10^{-2}M_p$) for slow-roll, DBI, and intermediate regimes. The bars on the graph indicate the range of tension possible: the largest range is for $m>10^{-5}M_p$.
\label{Gmuplot}}
\end{center}
\end{figure}

\section{Remarks and Summary}
We have examined how natural $D$3-$\D$3-brane inflation in string
theory is, and observational constraints on the parameter values. We find that the most basic condition of
inflation - sufficient e-folds - is easily met, so brane inflation is
quite natural in this sense. In fact, the sufficient e-folds condition
becomes easier to satisfy as the brane inflationary scenario becomes
more realistic in the context of string theory. Since the inflaton is an open string mode, and the inflaton potential
is generated by its one-loop effect (or in the cross channel, by the
exchange of a closed string) this property strongly guides its
potential. In addition, the DBI action goes a long way in allowing
enough e-folds of inflation. However, due to the many
compactification solutions in string theory, the inflationary
scenario will have a set of parameters. That is, due to the vast
cosmic landscape in string theory, the choice of vacua translates to
the choice of parameters. If $g_s$ and $m$ are fixed, only a narrow
range of choices for $V_0$ and $\lambda$ (narrower in slow-roll than
in DBI) will match the COBE normalization. But the two conditions of
enough e-folds and matching COBE data still allow for some
flexibility in the choice of other parameters, and varying these can
change the predictions for the observables (scalar index, tensor/scalar ratio, non-Gaussianities, cosmic string tension etc.).

Within this framework, we find observationally interesting cases where the non-Gaussianity or the tensor/scalar ratio $r$ saturates the present observational bound, and that
the power index $n_{s}$ covers the whole allowed range. Clearly,
better CMBR data will narrow the parameter ranges and help to
pinpoint the particular scenario in the early universe. As the data
improves, the analysis should be refined as well. In this paper, we
consider only a simplified warp factor, and we use 55 e-folds as a
canonical number. The warped deformed throat is quite well studied,
so there is plenty of room for improvement.
Our approach can be easily extended to incorporate other scenarios
for inflation, or additional terms in the potential. In any such
scenario, one should simply check that the parameters $\epsilon_D$, $
\eta_D$ and $\kappa_D$ stay less than 1 throughout 55 e-folds.
                                                                                
If the brane inflationary scenario is correct, this is a great probe
of the origin of our early universe and of the particular
compactification in string theory, telling us our location in the vast cosmic
landscape.

\vspace{0.3cm}

{\bf Acknowledgments}
We are grateful for enlightening discussions with Rachel Bean, Xingang Chen, Hassan Firouzjahi, Eanna Flanagan, Simon Gravel, Louis Leblond, Maxim Perelstein, Sash Sarangi, Ben Shlaer, Ira Wasserman and Mark Wyman. This work is supported by the National Science Foundation under grant PHY-0355005.

\vspace{0.3cm}

\vspace{0.3cm}

\section*{Appendix A: Potential}

We take the potential to be of the form 
\be
\label{potentialterms}V = V_K + V_A +V_{D\bar{D}}
\ee
where the last term is the usual Coulombic potential between the $D$3-${\D}$3-branes modified by the their relative velocity.

To calculate the velocity dependence, use the general expression for branes at angles\cite{Polchinski} for the case of Dp-branes with two non-zero angles: $\phi_1=\pi$ for the angle between branes and a complex angle $\phi_2=-iu$, where $tanh(u)=v$. Replacing the theta function in the denominator that is zero at $\phi_1=\pi$ by $(8\pi^2\ap t)^(1/2)e^{-\pi t}$, we find, for $D$3-${\D}$3-branes,
\be
V(r,v)=\frac{2}{(8\pi\ap)^2}\int_0^{\infty}dt\;t\;e^{-tr^2/(2\pi\ap)}e^{\pi t}\frac{\tanh(u)}{\eta(i/t)}\frac{[\theta_{11}(\frac{iu-\pi}{2\pi},\frac{i}{t})]^2[\theta_{11}(\frac{iu+\pi}{2\pi},\frac{i}{t})]^2}{\theta_{11}(\frac{iu}{\pi},\frac{i}{t})}
\ee
Taking the long distance limit, this simplifies to 
\be
V(r,v)=-\frac{(\gamma+1)^2}{4\gamma}\frac{1}{2\pi^3r^4}
\ee
As the branes move towards each other, the massive closed string modes start contributing to the potential. However, before their contribution becomes important, the tachyon mode appears, rapidly ending inflation (as in hybrid inflation) \cite{Sarangi:2003sg}. For this reason, we find the above potential 
$V(r,v)$ to be adequate for our purpose.

\vspace{0.3cm}

\bibliographystyle{JHEP}
\bibliography{Strings}

\end{document}